\documentclass{aa}

\usepackage{graphicx}
\usepackage{txfonts}
\usepackage{hyperref}
\usepackage{gensymb}
\usepackage{natbib}

\hypersetup{
	pdfnewwindow=true,  
	colorlinks=true,    
	linkcolor=blue,  
	citecolor=blue,    
	filecolor=cyan,     
	urlcolor=black,      
} 

\begin{document}

   \title{Diagnosing deceivingly cold dusty galaxies at $3.5<z<6$: A substantial population of compact starbursts with high infrared optical depths}

\titlerunning{Optically thick dust in galaxies at $3.5<z<6$}
\authorrunning{Jin et al.}

\author{
Shuowen Jin\inst{1,2,\thanks{Marie Curie Fellow}},
Emanuele Daddi\inst{3}, 
Georgios E. Magdis\inst{1,2,4},
Daizhong Liu\inst{5},
John R. Weaver\inst{1,4},
Qinghua Tan\inst{3,6},
Francesco Valentino\inst{1,4},
Yu Gao\inst{7},
Eva Schinnerer\inst{8},
Antonello Calabr\`o\inst{9},
Qiusheng Gu\inst{10,11},
David Blanquez Sese\inst{1,2}
          }

   \institute{Cosmic Dawn Center (DAWN)\\
      \email{shuji@space.dtu.dk}
    \and
             DTU-Space, Technical University of Denmark, Elektrovej 327, DK-2800 Kgs. Lyngby, Denmark
    \and             
            CEA, IRFU, DAp, AIM, Universit\'e Paris-Saclay, Universit\'e de Paris,
Sorbonne Paris Cit\'e, CNRS, F-91191 Gif-sur-Yvette, France
        \and 
    Niels Bohr Institute, University of Copenhagen, Jagtvej 128, 2200 Copenhagen, Denmark
    \and 
          Max-Planck-Institut f\"ur extraterrestrische Physik (MPE), Giessenbachstrasse 1, D-85748 Garching, Germany
      \and
    Purple Mountain Observatory, Chinese Academy of Sciences, 10 Yuanhua Road, Nanjing 210033, China
    \and 
    Department of Astronomy, Xiamen University, Xiamen, Fujian 361005, China
    \and 
    Max-Planck-Institut f\"ur Astronomie, K\"onigstuhl 17, D-69117 Heidelberg, Germany
     \and 
    INAF - Osservatorio Astronomico di Roma, via di Frascati 33, 00078, Monte Porzio Catone, Italy
     \and 
     School of Astronomy and Space Science, Nanjing University, Nanjing 210093, China
     \and 
     Key Laboratory of Modern Astronomy and Astrophysics, Nanjing University, Ministry of Education, China
             }

   \date{Received 16 February 2022 / Accepted 22 June 2022}

 \abstract
{
Using NOEMA and ALMA 3mm line scans, we measured spectroscopic redshifts of six new dusty galaxies at $3.5<z<4.2$ by solidly detecting [CI](1-0) and CO transitions. The sample was selected from the COSMOS and GOODS-North super-deblended catalogs with far-infrared (FIR) photometric redshifts $z_{\rm phot}>6$ based on template IR spectral energy distribution (SED) from known submillimeter galaxies at $z=$~4--6. Dust SED analyses explain the $z_{\rm phot}$ overestimate from seemingly cold dust temperatures ($T_{\rm d}$) and steep Rayleigh-Jeans (RJ) slopes, providing additional examples of cold dusty galaxies impacted by the cosmic microwave background (CMB). We therefore studied the general properties of the enlarged sample of 10 ``cold'' dusty galaxies over $3.5<z<6$.
We conclude that  these galaxies are deceivingly cold at the surface but are actually warm in their starbursting cores.
Several lines of evidence support this scenario: (1) The high infrared surface density $\Sigma_{\rm IR}$ and cold $T_{\rm d}$ from optically thin models appear to violate the Stefan-Boltzmann law;
(2) the gas masses derived from optically thin dust masses are inconsistent with estimates from dynamics and CI luminosities;
(3) the implied high star formation efficiencies would conflict with cold $T_{\rm d}$;  and (4)   high FIR optical depth is implied even using the lower, optically thick dust masses.
This work confirms the existence of a substantial population of deceivingly cold, compact dusty starburst galaxies at $z\gtrsim4$, together with the severe impact of the CMB on their RJ observables, paving the way for the diagnostics of optically thick dust in the early Universe. Conventional gas mass estimates based on RJ dust continuum luminosities implicitly assume an optically thin case, which leads to overestimation of gas masses by a factor of 2-3 on average in compact dusty star-forming galaxies.
}


   \maketitle


\section{Introduction}

The interstellar medium (ISM) is key to diagnosing the evolutionary state of galaxies in the early Universe. In particular, the available reservoir of molecular hydrogen ($M_{\rm H_2}$) dictates the ongoing star- and subsequent dust formation. However, measuring the molecular gas mass $M_{\rm gas}$ of galaxies, both in the local and the high$-z$ Universe is not a trivial task, as H$_2$ lacks dipole moment and thus is not detectable under the most typical, most representative ISM conditions. As gas and dust are well mixed (e.g., \citealt{Bohlin1978,Boulanger1996}), an efficient method to infer $M_{\rm gas}$ indirectly is to use far-infrared(FIR)/submillimeter(submm) dust continuum observations to derive the (observationally cheaper) dust mass ($M_{\rm dust}$) of a galaxy that is subsequently converted to $M_{\rm gas}$ through the metallicity-dependent dust-to-gas-mass ratio technique (e.g., \citealt{Leroy2011,Eales2012,Magdis2012SED,Magdis2017,Santini2014,Genzel2015,Berta2016,Schinnerer2016,Tacconi2018,Tacconi2020,Liu_DZ2019b,Kokorev2021}). Similarly,  single band measurements of the dust emission flux on the Rayleigh-Jeans side of the SED have also been calibrated against CO observations and under a series of assumptions (e.g., dust temperature, $\alpha_{\rm CO}$ conversion factor) can be used to infer gas mass estimates (e.g., \citealt{Groves2015,Scoville2016,Scoville2017ISM,Schinnerer2016,Liu_DZ2019b}).

However, the derivation of $M_{\rm gas}$ from $M_{\rm dust}$ or from a single band Rayleigh-Jeans (RJ) dust continuum flux density are  usually based on the assumption that the observed FIR/submm emission is optically thin. If instead the dust opacity of a galaxy extends to $\lambda>100\mu$m, its $M_{\rm dust}$ based on optically thin models can be severely overestimated (a factor of 2--20 e.g., \citealt{Galliano2011,Jin2019alma,Cortzen2020GN20,Fanciullo2020}). Subsequently, the inferred $M_{\rm gas}$ would also be problematic. While there is mounting evidence for optically thick dust emission in both local and high-z dusty star-forming galaxies (e.g., \citealt{Blain2003,Draine2007SED,Conley2011,Casey2012SED,Spilker2016,Hodge2016,Scoville2017Arp220,Simpson2017,Riechers2014AzTEC3,Riechers2017,Riechers2020,Cortzen2020GN20}), the large degeneracy between the solutions of the thick and thin dust model SEDs prevents a definitive conclusion.

Besides the dust SEDs, submm lines provide more powerful diagnostics for ISM properties, revealing the presence of optically thick dust emission at FIR wavelengths in the starbursting cores of submillimeter galaxies (SMGs) in the local Universe. For example, \cite{Papadopoulos2010Arp220} developed a method to identify the optically thick dust using ratios between optically thick and optically thin lines in local galaxies (e.g., HCN(1-0)/CO(1-0) and CO(6-5)/CO(3-2)) and proposed that dust optical depths can significantly suppress high-J CO and C$^+$ emission. Based on James Clerk Maxwell Telescope (JCMT) observations, they found that the very faint CO(6-5) lines in Arp220 and other ultraluminous infrared galaxies (ULIRGs) can be attributed to high dust optical depths at FIR wavelengths immersing those lines in a strong dust continuum.
Furthermore, using high-resolution ALMA CO(1-0) mapping, \cite{Scoville2017Arp220} confirmed that the dust in the two nuclei of Arp 220 is optically thick well into the FIR, and in one of them even at 2.6\,mm with a high dust brightness temperature of 147~K.

Such high optical depths are also expected in the ISM of high-z star-forming galaxies that were more compact and rich in gas and dust. For example, \cite{Jin2019alma} discovered a sample of ``apparently'' extremely cold dusty galaxies at $z=4$--6 using ALMA line scans. Their dust spectral energy distributions (SEDs) peak at longer rest-frame wavelengths than most literature sources at $z>4$, and their dust temperatures (24--42~K) as inferred with optically thin models are between 1.5 and 2.0 times colder than that of main sequence (MS) galaxies at these redshifts. Given the massive dust content and compact morphology of these galaxies, the ``apparent'' cold dust temperatures are best understood as evidence for large optical depths that effectively attenuate the dust continuum emission in the FIR and shift the peak of the observed SED to longer wavelengths. 
In the same direction, \cite{Cortzen2020GN20} used the two neutral carbon transitions in GN20, a star-bursting galaxy at $z=4.05$, and found that the excitation temperature as inferred by the [CI](2-1)/(1-0) ratio is significantly higher than the abnormally cold dust temperature derived from SEDs under the assumption of an optically thin FIR dust emission. On the other hand, the excitation temperature is fully consistent with the dust temperature of a general opacity model with $\lambda_{\rm eff} \approx 170\mu$m. These findings suggest that high FIR optical depths could be a widespread feature of starbursts across comic time. Follow-up observations of seemingly ``cold'', high-$z$ star-bursting galaxies with compact-dust-emitting sizes are necessary to push this investigation forward. 

In this paper, we present an in-depth investigation of the ISM properties of six such galaxies based on new NOEMA and ALMA observations. These six galaxies are used to extend the original sample of  \cite{Jin2019alma} to a total of ten, which is used in the remainder of this paper. In Section 2 we describe the sample selection and data processing. In Section 3 we discuss redshift confirmation, SED modeling, and size measurements. We present evidence of optically thick dust in Section 4. We discuss the implication of these results in Section 5 and provide our summary and conclusions in Section 6.
We adopt cosmology $H_0=73$, $\Omega_M=0.27$, $\Lambda_0=0.73$; and a Chabrier initial mass function (IMF) \citep{Chabrier2003}.

\section{Sample, observations, and data reduction}

\begin{table*}
{
\caption{Overview of new observations presented in this paper}
\label{tab:1}
\centering
\begin{tabular}{cccccccccccc}
\hline
\hline
     ID$^*$       & R.A., Dec. &  3mm   & Beam$_{\rm 3mm}$ & 1mm  &  870$\mu$m &   line scan\\   
                  \hline
3117 & 150.58858, 2.31908 & S18DI & $3.4''\times2.3''$ & --  & 2016.1.00463.S & NOEMA\\
5340 & 150.34488, 2.01271 & S18DI,W20DM &  $5.1''\times3.9''$ & --  & 2016.1.00463.S & NOEMA\\
6309 & 150.08060, 1.68511 & S18DI,W20DM &  $5.7''\times4.3''$ & 2016.1.00279.S & 2016.1.00463.S  & NOEMA \\
7549 & 149.54492, 2.10318 &  S18DI &  $4.0''\times1.9''$ & 2016.1.00279.S  & 2016.1.00463.S  & NOEMA \\
9316 &  149.99498, 2.58278 & 2018.1.00874.S &  $0.41''\times 0.36''$ & 2013.1.00118.S & 2016.1.00478.S & ALMA\\ 
12646$^*$ & 189.11469, 62.20501 &  W17EK,W19DQ & $1.6''\times 1.1''$ & W18EY   
& -- & NOEMA\\
\hline
\end{tabular}\\
}
{Notes: $^*$ID12646 is from the \cite{Liu2018} catalog, while the remaining IDs are originally  from the radio catalog of \cite{Smolcic2017} and are also used in the COSMOS super-deblended catalog (ID$_{\rm Jin} =$ID$_{\rm Smolcic}$+2E7).}
\end{table*}

\begin{figure*}
\centering
\includegraphics[width=0.98\textwidth,trim={0cm 0cm 0cm 0cm}, clip]{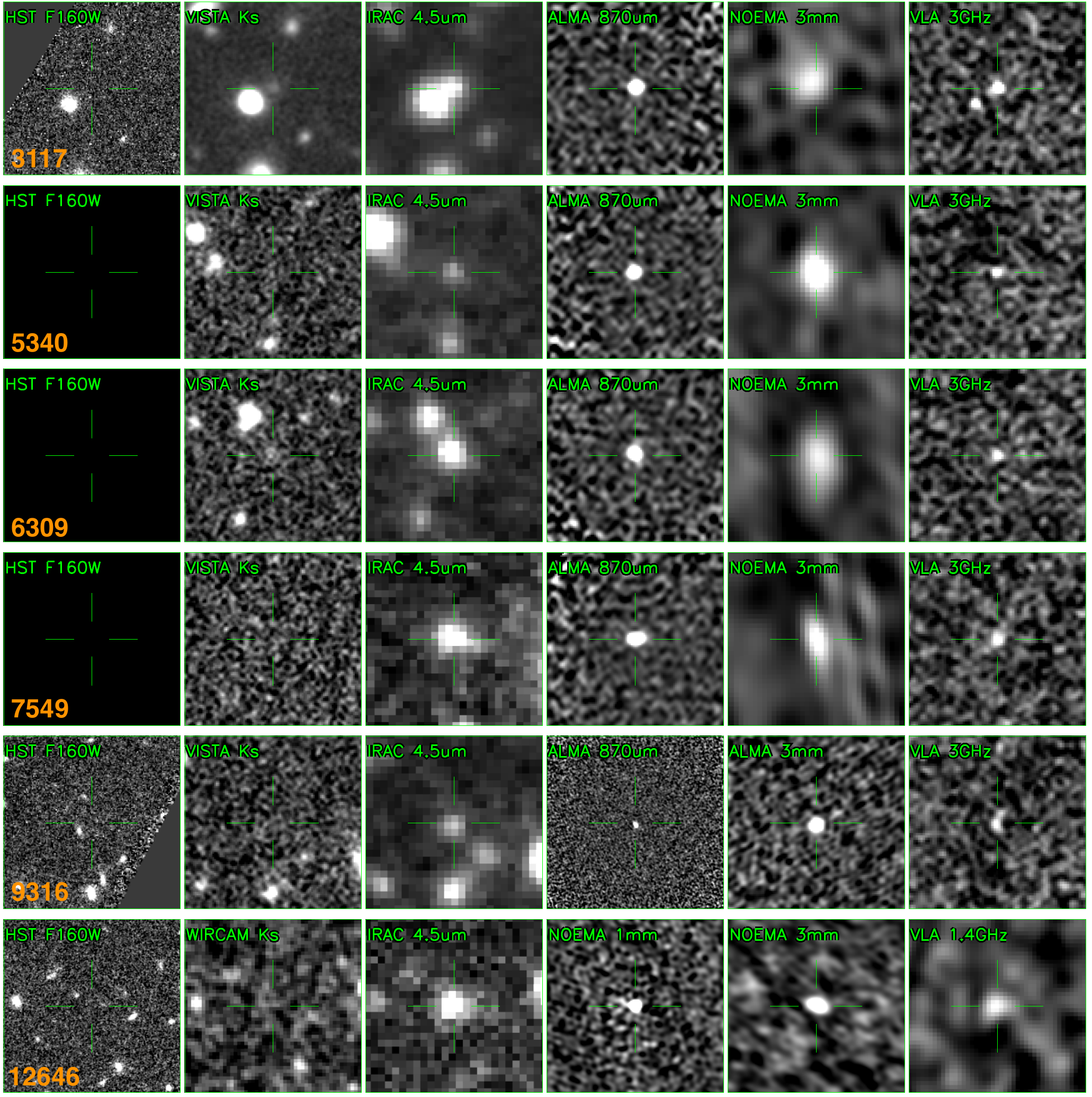}
\caption{
Multi-wavelength images with $15''\times15''$ sizes. The instrument
used and wavelength are shown in green text in each cutout. ID5340, 6309, and 7549 have no HST F160W data available and thus are shown as blank panels.
\label{cutout}
}
\end{figure*}

Six new dusty star-forming galaxies were selected for follow-up observations in this study, after being culled for potentially lying at very high redshifts ($z>6$) based on their SEDs.
We show their multi-wavelength images in Fig.~\ref{cutout}, including deep ALMA and NOEMA continuum data. 
The NOEMA and ALMA observations are summarized in Table~\ref{tab:1}. Five are in the COSMOS field and one is in the GOODS-North field. 

Among the five COMSOS sources, ID3117, 5340, 6309, and 7549 were selected from the COSMOS super-deblended catalog \citep{Jin2018cosmos}. They were originally detected in the VLA 3GHz catalog \citep{Smolcic2017} and are optically dark, that is, with no detection in HST $I$ or $H$ band, and are therefore not included in the COSMOS2015 catalog \citep{Laigle2016}.
In the COSMOS super-deblended catalog \citep{Jin2018cosmos}, these four galaxies are detected with significant FIR emission (S/N$_{\rm FIR+mm}>$ 5) in Herschel, SCUBA2, AzTEC, and/or MAMBO \citep{Geach2016,Cowie_2017,Aretxaga2011,Bertoldi2007}. 
Fitting of their dust SEDs returns very high photometric redshifts of $z_{\rm phot,FIR}> 6$ using the $z=6.3$ HFLS3 template \citep{Riechers2013Nature}. We carried out follow-up observations of the four galaxies with NOEMA 3mm line scans (ID: S18DI, PI: S. Jin) in the winter 2018 semester. 

ID9316 is a bright submm galaxy close to a galaxy group RO0959 at $z=3.1$ \citep{Daddi2022clusters}, and was proposed to be at $z>7$ by Oteo and collaborators (see ALMA proposal ID: 2018.1.00874.S, PI: I. Oteo) and was observed with ALMA standard line scans in band 3. Meanwhile, this galaxy has also snapshot imaging in ALMA band 6 (ID: 2013.1.00118.S, PI: M. Aravena) and band 7 (ID: 2016.1.00478.S, PI: O. Miettinen).

ID12646 was selected from the FIR/(sub)mm super-deblended catalog in the GOODS-North field \citep{Liu2018} as the source with the highest photometric redshift $z_{\rm phot,FIR}\sim6$. During 2017 to 2019, Liu et al. carried out follow-up analyses of this source with NOEMA 3mm (IDs: W17EK, W19DQ) and 1mm observations (ID: W18EY).



\subsection{NOEMA}

The four galaxies, ID3117, 5340, 6309, and 7549, were observed with NOEMA 3mm line scans in the winter of  2018 (ID: S18DI, PI: S. Jin), continuously covering the frequency range 74.7--106~GHz. The observations were conducted in track-sharing mode with array configurations C and D. Each source was observed with two to three tracks in each tuning, reaching rms sensitivities of 0.13--0.16 mJy per 500 km/s with synthesized beams as listed in Table 1.
We re-observed the two sources ID5340 and 6309 with one NOEMA  tuning setup in January and February 2021 (ID: W20DM, PIs: E. Daddi \& S. Jin) in order to obtain solid [CI](1-0) detections. ID7549 is also well-detected at 2mm continuum with NOEMA (ID: W20DM, PI: S. Jin).

ID12646 was firstly followed up with NOEMA 3mm line scans in the winter of  2017 (IDs: W17EK, PI: D. Liu); however, only one line was detected at 89.7~GHz. As the line might have been  CO(6-5) at $z=6.7$, a NOEMA 1mm observation (ID: W18EY, PI: D. Liu) was carried out to search for the [CII]158um line, but no line was detected. In the winter of  2019, this galaxy was observed with a NOEMA 3mm setup (ID: W19DQ, PI: D. Liu).

The NOEMA data were reduced and calibrated using IRAM GILDAS software packages. We then produced $uv$ tables to perform further analysis with IRAM GILDAS working on the $uv$ space (visibility) data.
The resulting synthesized beam at 3mm is in most cases rather elongated (Table 1) and no source is expected to be resolved given the resolution from the compact NOEMA configurations used. We therefore extracted the spectra by fitting a point source model in the $uv$ space at the known spatial positions from the high-resolution ALMA imaging. 
The NOEMA spectra are shown in Fig.~\ref{spectra}.

\subsection{ALMA}

The above-mentioned four NOEMA sources in COSMOS have public imaging data in the ALMA archive at 345~GHz (ID: 2016.1.00463.S, PI: Y. Matsuda),  in which two of them are also observed at 210~GHz  (ID: 2016.1.00279.S; PI: I. Oteo). The four galaxies are solidly detected in  the dust continuum and their fluxes were accurately measured. No serendipitous line was found in the ALMA data cubes.

ID9316 was observed with five ALMA setups at 3mm, continuously covering frequencies from 89.5 to 103.5~GHz.
We processed each setup by reproducing the observatory calibration with their custom-made script based on the Common Astronomy Software Application package (CASA; \citealt{McMullin2007}). We then exported the data into uvfits format to generate $uv$ tables to perform further analysis with the IRAM GILDAS working on the $uv$ space (visibility).
The spectrum of each setup is extracted in  the $uv$ space on the position of the continuum peak using a Gaussian kernel determined from the continuum and kept fixed in size and position in each spectral channel.
The final spectra, as shown in Fig.~2, are produced by stacking the spectra of the five setups. We measure a redshift of $z=4.072$ via detecting multiple lines, including CO(4-3), [CI](1-0), and CO(5-4).
The 1mm and 870$\mu$m ALMA data of ID9316 were well processed as part of the A$^3$COSMOS catalog \citep{Liu2019A3COSMOS}. We therefore directly adopt their measurements.

As a sanity check, we compared the ALMA 870$\mu$m photometry for all COSMOS sources with the SCUBA2 850$\mu$m photometry from the super-deblended catalog, finding that they are in excellent agreement (see a similar comparison, with consistent results, in \citealt{Jin2018cosmos} Fig. 18 and 19). The source ID3117 has ACA (Atacama Compact Array) 630$\mu$m observations (2019.1.01832.S, PI: J. Zavala), and its 630$\mu$m flux from our SED fitting also agrees well with the ACA photometry. Together, these checks confirm the expected high fidelity of the flux measurements from the super-deblended technique.

%

We emphasize that we analyzed  both NOEMA and ALMA data in $uv$ space using the same algorithms that were applied in \cite{Jin2019alma}, \cite{Puglisi2019}, and \cite{Valentino2018CI,Valentino2020CO,Valentino2020CI}. 
FIR/(sub)mm photometry is measured using the same Super-deblending technique \citep{Jin2018cosmos,Liu2018}. The consistencies therefore allow us to directly compare results in this work to these previous studies and to merge the samples.

\section{Results}

\subsection{Line detection and redshift identification}

\begin{figure*}
\setlength{\abovecaptionskip}{-0.1cm}
\setlength{\belowcaptionskip}{-0.4cm}
\centering
\includegraphics[width=0.95\textwidth]{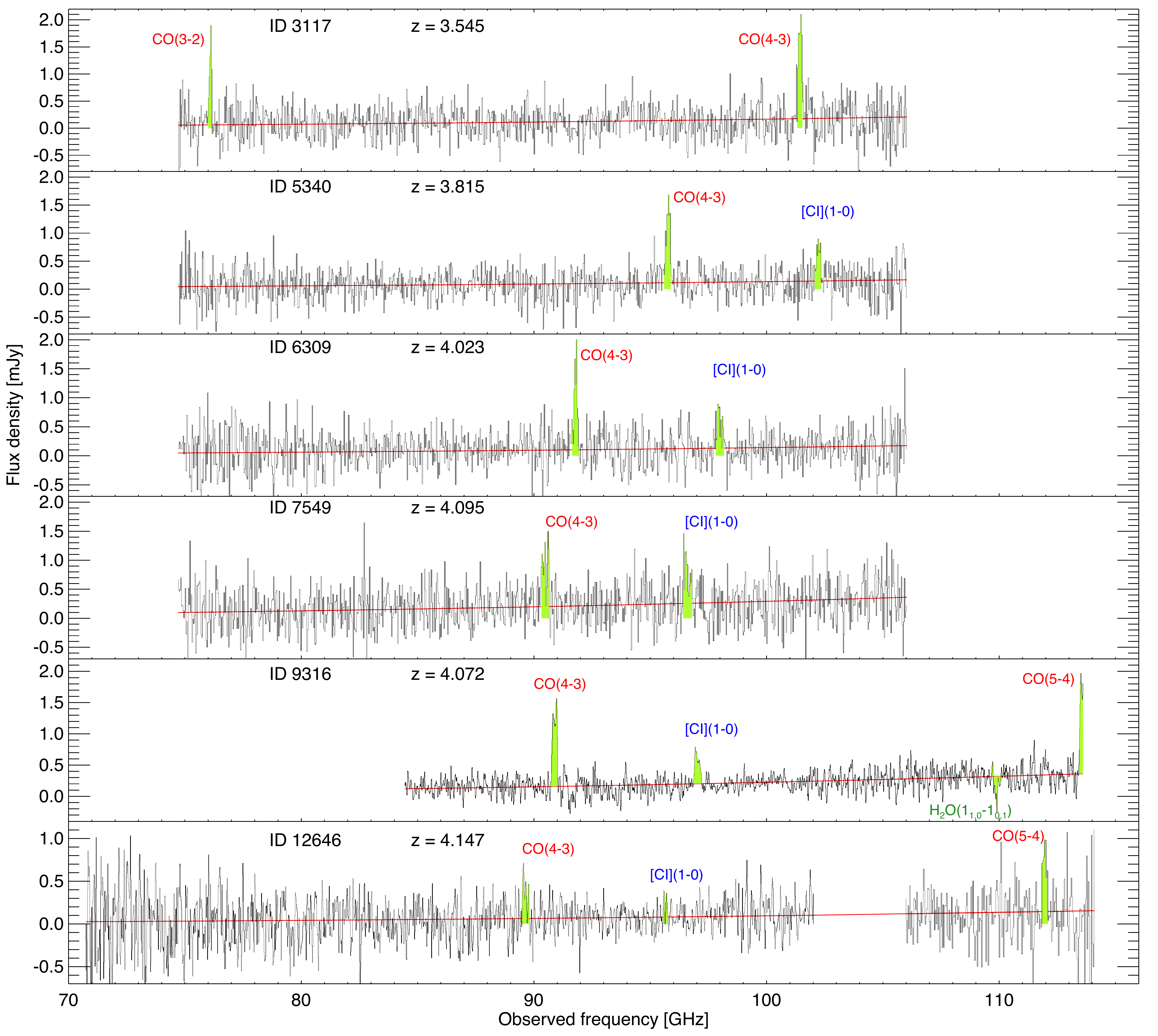}
\caption{%
NOEMA and ALMA (ID9316) 3mm spectra. Detected lines are highlighted in green with their names labeled. The red line shows a power-law fit to the continuum that increases with frequency $\nu^{3.7}$.
\label{spectra}
}
\end{figure*}

Adopting the line-search algorithm from \cite{Jin2019alma}, we blindly searched for lines by sweeping through the full 3 mm continuum-subtracted spectrum.
We present their NOEMA and ALMA 3mm spectra in Fig.~\ref{spectra} with detected transitions highlighted in green, and the line measurements are listed in Table~\ref{tab:2}. Every galaxy is detected in at least two transitions, where the brightest lines are detected with 6--14$\sigma$ and secondarily bright ones are detected at 4.5--6.0$\sigma$.
We thus confirm redshifts of all sources to be at $z=$3.545--4.147. 
ID3117 has the lowest redshift in this sample, of namely $z=3.545$, which is confirmed by CO(3-2) and CO(4-3) emission lines (S/N=9, 13).
ID5340, 6309, and 7549 are confirmed at $z=3.815, 4.032,$ and 4.095 with CO(4-3) and [CI](1-0) emission lines, respectively. 
ID12646 is detected with three lines: CO(5-4), CO(4-3), and [CI](1-0) at $z=4.147$.
ID9316 is confirmed at $z=4.074$ with ALMA, detecting CO(4-3), CO(5-4), and [CI](1-0) emission with S/N=30, 23, and 19, respectively. An absorption of H$_2$O(1$_{10}$-1$_{01}$) is also detected with S/N=6.2, which appears two times stronger than the continuum at the same frequency, similar to the strong H$_2$O absorption found in HFLS3 by \cite{Riechers2022CMB}.
Notably, five of the six sources are detected with [CI](1-0) transition at 4.5--19$\sigma$. Given that most CI detections are in the local Universe (e.g., \citealt{Jiao2017CI,Jiao2019CI,Jiao2021CI}), the largest high-z [CI] sample is at $z\sim1.2$ from \citep{Valentino2018CI,Valentino2020CI}, and several [CI](2-1) detections at $z\sim3$ were presented by \cite{Yang2017CI21}. This [CI] sample is comparable in size to the SPT ones \citep{Vieira2013,Bothwell2017CI}, thus constituting one of the first sizable [CI] samples  at $z\gtrsim4$.

Remarkably, the redshifts of the five NOEMA sources are all lower than the original photometric redshifts. 
In the same way as the four galaxies found by \cite{Jin2019alma}, this sample was selected as photometric $z>6$ candidates but eventually confirmed at $z\sim4 $.
We note that the redshift of ID9316 is consistent with our own $z_{\rm phot,IR}=3.9\pm 0.8$ from the super-deblended catalog \citep{Jin2018cosmos}, but we keep this galaxy in the sample for completeness, distinguishing it from the remaining galaxies where appropriate. The full sample of ten galaxies is studied here as an enlarged version of the sample studied by \cite{Jin2019alma}.

\begin{table*}
{
\caption{Line detections}
\label{tab:2}
\centering
\begin{tabular}{ccccccccccccc}
\hline
\hline
     ID       &  $z_{\rm spec}$   & $I_{\rm CO(4-3)}$ & $I_{\rm [CI](1-0)}$ & $I_{\rm CO(3-2)}$&  $I_{\rm CO(5-4)}$  & $I_{\rm H{_2}O(1_{1,0}-1_{0,1})}$ & FWZI \\
             &          &        [Jy km/s]        &          [Jy km/s]                  &     [Jy km/s]        &   [Jy km/s]   &   [Jy km/s]   &   [km/s]           &  \\                 
                  \hline
3117 & 3.545    &  0.90$\pm$0.07   & --  & 0.71$\pm$0.08  &    -- & -- &   568  \\
5340 & 3.815    &    0.90$\pm$0.06 & 0.35$\pm$0.06    & --  &    --  & -- &   777   \\
6309 & 4.023     &    0.65$\pm$0.07 &  0.46$\pm$0.10  & -- &    --  & --  &   444    \\
7549 &  4.095    &  0.81$\pm$0.12 & 0.60$\pm$0.12    & -- &   -- &  -- &  1008     \\
9316 &  4.072     &    0.89$\pm$0.03 &    0.36$\pm$0.02    & --  &    1.16$\pm$0.05$^a$ & 0.14$\pm$0.02$^{b}$ &  1056 \\ 
12646 &  4.147     &    0.48$\pm$0.08 &    0.24$\pm$0.05   & --  &    0.78$\pm$0.11 & -- & 700 \\
\hline
\end{tabular}\\
}
{Notes: $^a$The CO(5-4) line is on edge of the spectra, we estimated the line flux assuming an identical line profile to CO(4-3). $^b$ The ${\rm H{_2}O(1_{1,0}-1_{0,1})}$ line is in absorption.}
\end{table*}

\begin{figure*}
\centering
\includegraphics[width=0.9\textwidth]{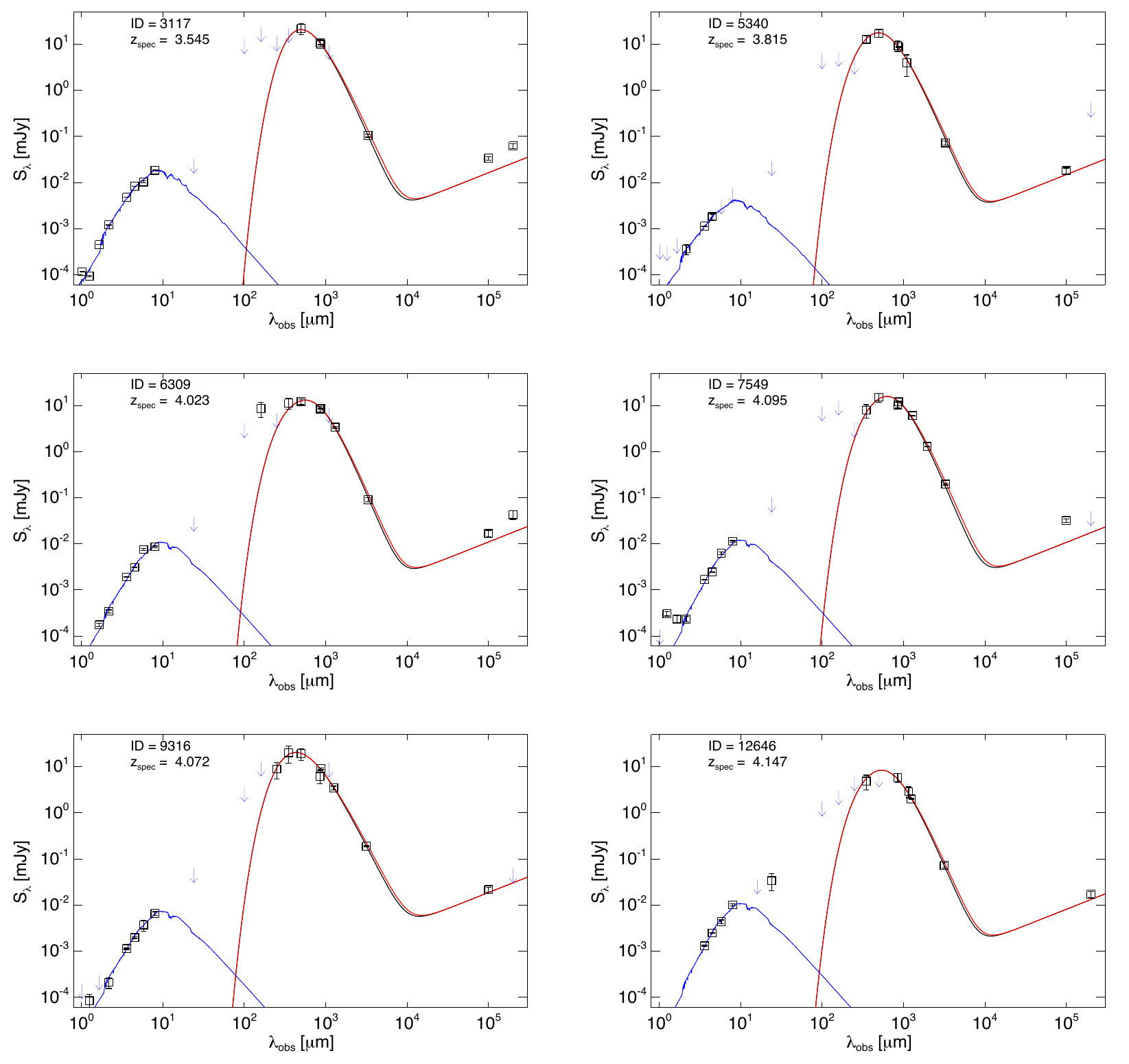}
\caption{
SEDs of galaxies in this study. Photometry is from the super-deblended catalogs \citep{Jin2018cosmos,Liu2018} and ALMA and NOEMA measurements. The dust SEDs are fitted with optically thick MBB models accounting for CMB impact on the continuum \citep{daCunha2013CMB}. The black (red) curve shows the best fit to the observed SED accounting (not accounting) for CMB. The blue curve marks the stellar component. We note that the radio data are not included in the fitting; we extrapolated a radio component to the dust SEDs using the IR--radio correlation from \cite{Delvecchio2021}.
        }
\label{sed}
\end{figure*}

\begin{table}[ht!]
{
\caption{NOEMA and ALMA continuum measurements}
\label{tab:3}
\centering
\begin{tabular}{cccc}
\hline
\hline
     ID       &  $S_{\rm 3mm}$ [mJy]&  $S_{\rm 1.2mm}$~$^{(c)}$ [mJy] &   $S_{\rm 870\mu m}$$^{(d)}$ [mJy] \\
3117   & 0.104$\pm$0.008$^{(a)}$& --  & 9.85$\pm$0.17 \\
5340  &  0.072$\pm$0.008$^{(a)}$  & -- & 8.63$\pm$0.17 \\
6309  & 0.090$\pm$0.008$^{(a)}$  & 3.37$\pm$0.14 &  8.62$\pm$0.19 \\
7549  & 0.196$\pm$0.008$^{(a)}$ &   6.03$\pm$0.12 & 11.92$\pm$0.19 \\
9316  & 0.188$\pm$0.004$^{(b)}$  & 3.47$\pm$0.31 & 8.99$\pm$0.42  \\
12646  & 0.073$\pm$0.010$^{(b)}$ & 2.00$\pm$0.054 &--\\
\hline
\end{tabular}\\
}
{Notes. The dust continua are measured at observed frequencies of (a) 90.4~GHz; (b) 95~GHz; (c) 240~GHz, and (d) 345~GHz.}
\end{table}

\begin{table*}
{
\caption{CMB impact on observables}
\label{tab:4}
\centering
\begin{tabular}{cccccccccccc}
\hline\hline
     ID       &   $L_{\rm IR}$  &    $\beta_{\rm thin}^{\rm noCMB}$ &   $\beta_{\rm thin}^{\rm CMB}$ &  $\beta_{\rm thick}^{\rm CMB}$ &   $T_{\rm d,thin}^{\rm noCMB}$ &   $T_{\rm d,thin}^{\rm CMB}$ &   $T_{\rm d,thick}^{\rm CMB}$  & $M_{\rm d,thin}^{\rm noCMB}$ & $M_{\rm d,thin}^{\rm CMB}$ & $M_{\rm d,thick}^{\rm CMB}$ & $\lambda_{\rm eff}$  \\
                  &    [$10^{12}L_\odot$]      &                &             &       &  [K]          &  [K]   & [K]       &  [$10^{8}$M$_\odot$] & [$10^{8}$M$_\odot$]  & [$10^{8}$M$_\odot$]  & [$\mu$m]\\
 \hline
 (Cold)\\
3117    &    6.24$\pm$2.62  &  2.5$\pm$0.2 &    2.3$\pm$0.2   & 2.2$\pm$0.2 &      25$\pm$4     &   26$\pm$4 & 30$\pm$9  &   20.0$\pm$4.0 &  26.5$\pm$7.8 & 21.2$\pm$9.1 & 100$\pm$23 \\
5340    &      5.85$\pm$0.81  &    2.7$\pm$0.2 &    2.4$\pm$0.1 & 2.2$\pm$0.2   &      25$\pm$1     &   26$\pm$1 &  41$\pm$10  &  12.2$\pm$1.1 & 17.9$\pm$2.3  & 8.0$\pm$3.7 & 150$\pm$30\\
6309    &   4.82$\pm$0.84  &     2.6$\pm$0.2  &    2.3$\pm$0.2   &   2.3$\pm$ 0.3&    24$\pm$2     &   25$\pm$2 &    42$\pm$6 & 16.3$\pm$1.6  & 23.5$\pm$3.5 & 8.7$\pm$2.9 & 200$\pm$19\\
7549    &   4.90$\pm$0.87  &    2.6$\pm$0.2 &    2.1$\pm$0.2  & 2.1$\pm$0.2  &      22$\pm$2     &   24$\pm$3 &   36$\pm$7  & 38.2$\pm$4.3  & 61.6$\pm$1.6 & 25.7$\pm$5.8  & 200$\pm$23\\
12646    &   3.50$\pm$0.61 &  2.5$\pm$0.2 &    2.2$\pm$0.2  & 2.2$\pm$0.2 &      26$\pm$2     &   27$\pm$2 &   43$\pm$9  & 8.8$\pm$0.9    & 12.9$\pm$1.8 & 6.1$\pm$2.5  & 175$\pm$24\\
\hline
(Warm)\\
9316    &   10.01$\pm$2.48 &   1.8$\pm$0.1 &    1.6$\pm$0.1  & 1.6$\pm$0.2  &   36$\pm$3 &    37$\pm$3  & 46$\pm$8  & 19.0$\pm$2.2 & 23.1$\pm$3.4 & 16.5$\pm$4.4  & 100$\pm$28 \\
\hline
\hline
\end{tabular}\\
}
{Notes:   
$L_{\rm IR}$: IR luminosity at 8--1000${\rm \mu m}$ rest-frame corrected for CMB effect; 
$\beta_{\rm thin}^{\rm noCMB}$, $T_{\rm d,thin}^{\rm noCMB}$ and $M_{\rm d,thin}^{\rm noCMB}$: Rayleigh-Jeans slope, dust temperature, and dust mass from optically thin MBB fits without accounting for CMB; 
$\beta_{\rm thin}^{\rm CMB}$, $T_{\rm d,thin}^{\rm CMB}$, and $M_{\rm d,thin}^{\rm CMB}$: Rayleigh-Jeans slope, dust temperature, and dust mass from optically thin MBB fits accounting for CMB effect; 
$\beta_{\rm thick}^{\rm CMB}$, $T_{\rm d,thick}^{\rm CMB}$, and $M_{\rm d,thick}^{\rm CMB}$: Rayleigh-Jeans slope, dust temperature, and dust mass from optically thick MBB fits accounting for CMB effect;
$\lambda_{\rm eff}$: effective wavelength of optical depth $\tau=1$.
}
\end{table*}

\begin{figure*}
\setlength{\abovecaptionskip}{-0.1cm}
\setlength{\belowcaptionskip}{-0.4cm}
\centering
\includegraphics[width=0.98\textwidth,trim={0cm 0cm 0cm 0cm}, clip]{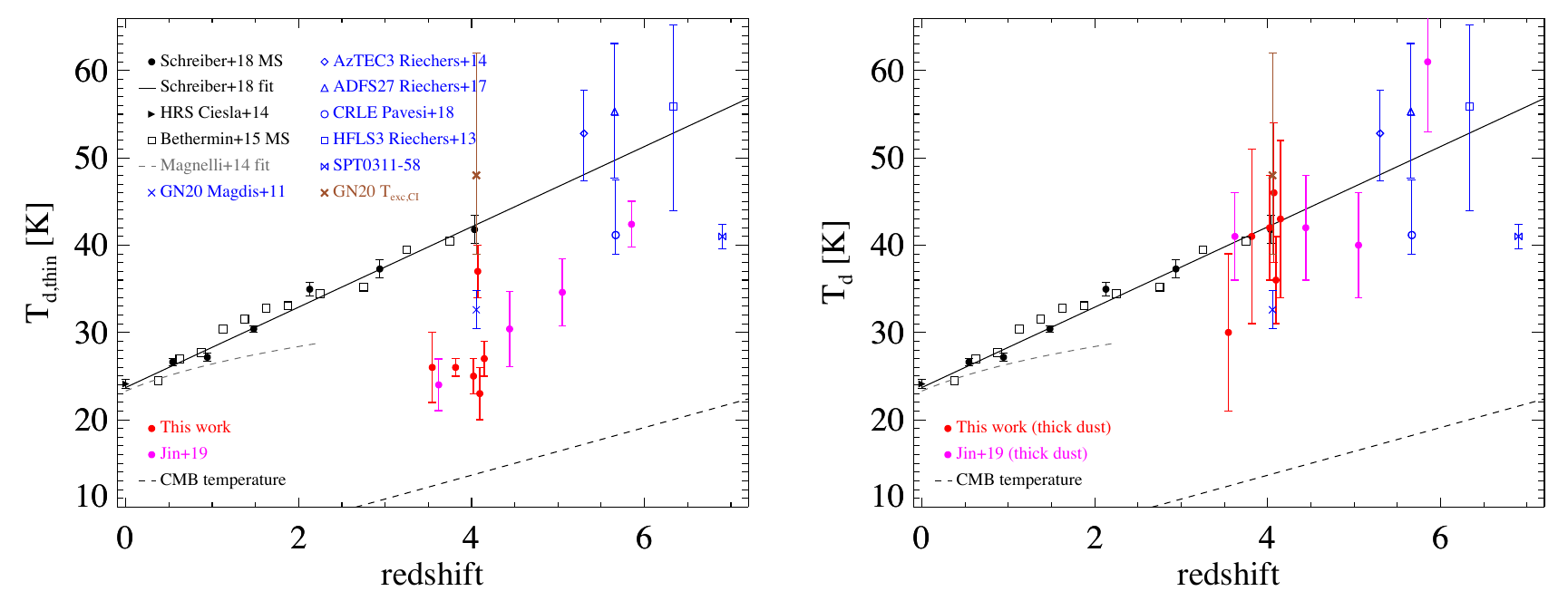}
\caption{
Dust temperature $T_{\rm d}$ versus redshift for galaxies in the literature and this work. 
Left panel shows dust temperatures in optically thin models. Right panel shows dust temperatures in optically thick models for galaxies in \cite{Jin2019alma} and this work, and in optically thin $T_{\rm d}$ for remaining data points. 
This figure shows a population of cold dusty galaxies at $z=$4--6 in optically thin models, but they might be comparably warm, as in main sequence galaxies, if they have optically thick dust.
\label{Td}
}
\end{figure*}

\subsection{Cold dust SEDs and CMB impact}

As mentioned above, the sources are confirmed at lower redshifts than the early expectations. To estimate the characteristic dust temperatures, we performed a detailed SED analysis by fitting their deblended FIR photometry and ALMA/NOEMA continuum fluxes.
Following the identical SED recipes adopted in \cite{Jin2019alma}, we fitted the FIR/(sub)mm photometry of this sample in three ways, using: (1) optically thin MBB (modified black body) models from \cite{Magdis2012SED}; (2) optically thin MBB models from \cite{Magdis2012SED} accounting for the CMB impact on the dust continuum following \cite{daCunha2013CMB}; and (3) optically thick MBB models from \cite{Magdis2012SED} also including  CMB following \cite{daCunha2013CMB}. The results are listed in Table~\ref{tab:4} and optically thick SEDs are presented in Fig~\ref{sed}. 
The radio data points are not directly used for the fittings, but  we extrapolated the expected radio emission based on the measured IR luminosities  using the mass-stratified IR--radio correlation from \cite{Delvecchio2021}. The predicted radio fluxes are in good agreement with the observations for most cases in this sample, showing little evidence for radio AGN emission.

The five NOEMA sources show clearly steep Rayleigh-Jeans slopes $\beta>2.5-2.7$ for optically thin MBB models; see the $\beta_{\rm thin}^{\rm noCMB}$ listed in Table~\ref{tab:4}. These Rayleigh-Jeans slopes are similar and even steeper than the findings in \cite{Jin2019alma}. Benefiting from high signal-to-noise ratios (S/Ns) from the ALMA and NOEMA photometry, the constraints on the Rayleigh-Jeans slopes are more robust than those in \cite{Jin2019alma}.
However, no indisputable evidence has yet been produced in the literature for galaxies with steep Rayleigh-Jeans slope $\beta>2$. A recent ALMA large survey revealed that the ALESS SMG sample can be characterised by $\beta\simeq 1.9\pm0.4$ \citep{daCunha2021beta}, which is consistent with the emissivity index of the dust in the Milky Way and other local and high-z galaxies, indicating little evolution in dust grain properties between high-z SMGs and local dusty galaxies. 
Therefore, the steep Rayleigh-Jeans slopes $\beta>2$ appear not to be genuine in these galaxies, similarly to what is discussed in \cite{Jin2019alma}.
On the other hand, the CMB temperature is warmer at higher redshifts, reaching 14~K at $z=4$.  This reduces the observable dust continuum of cold dusty galaxies notably at the longest wavelengths \citep{daCunha2013CMB,Zhang2016CMB}; e.g., 3mm. This suggests that the CMB has an impact on the Rayleigh-Jeans slopes and accounting for it can explain their abnormally steep observed values \citep{Jin2019alma}.

After accounting for the CMB effect on the continuum using \cite{daCunha2013CMB} models, the Rayleigh-Jeans slopes reduce to lower best-fitting values $\beta=$2.0--2.5.
We note that the best-fitting Rayleigh-Jeans slopes of ID3117, 5340, and 6309 remain at $\beta>2$ in optically thin models even though the CMB effect has been accounted for. Their slopes can be further reduced to $\beta\sim2$ if we fit their SEDs using an optically thick dust template (plus the CMB effect).

In Fig.~\ref{Td}, we compare the dust temperatures from optically thin models to literature data. In the five NOEMA sources, their dust temperatures from optically thin models are 24--26 K, which is two times colder than the average $T_{\rm d}$ of main-sequence (MS) galaxies at the same redshift \citep{Schreiber2018Td}, and similar to the extreme cold sample in \cite{Jin2019alma}.
This work therefore confirms the existence of a new population of apparently cold dusty galaxies at $z\approx4$, further strengthening the conclusion drawn by \cite{Jin2019alma}.
However, their dust temperatures would be two times warmer if the dust is optically thick in FIR (Table~\ref{tab:4}), thus becoming comparably warm as typical dusty star-forming galaxies at high-z (Fig.~\ref{Td}-right). 
We present a diagnosis of the dust opacity of this sample in Sect. 4.

\subsection{Size measurements}
\begin{figure}
\setlength{\abovecaptionskip}{-0.1cm}
\setlength{\belowcaptionskip}{-0.4cm}
\centering
\includegraphics[width=0.48\textwidth,trim={2cm 0.5cm 0cm 0cm}, clip]{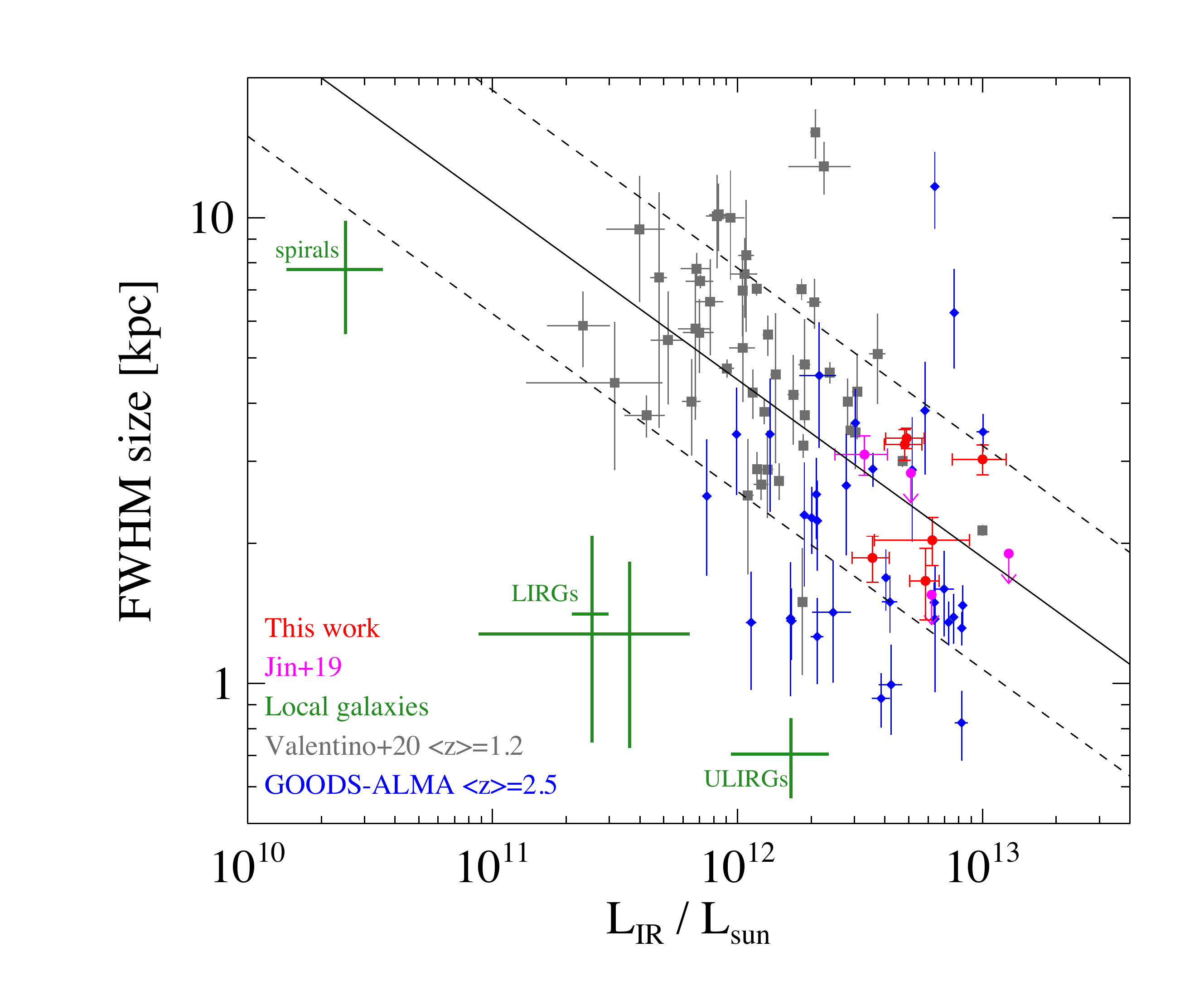}
\caption{
Size of dust continuum vs IR luminosity for this sample and data in the literature. Grey diamonds represent MS galaxies at $z\sim1.2$ in \cite{Valentino2020CI} and blue squares mark the GOODS-ALMA sample \citep{Franco2020alma,Gomez-Guijarro2022}. {Green crosses show the medians of four local galaxy samples (Spirals: \citealt{Bolatto2017}; LIRGs: \citealt{Bellocchi2022size}; ULIRGs: \citealt{Pereira-Santaella2021}); the dispersion of each sample is indicated by an error bar}. The solid line shows the best fit to {the high-z} data points, and the dashed lines mark the 1$\sigma$ scatter. Both local and high-z galaxies show an anti-correlation between dust continuum size and infrared luminosity.
\label{size_LIR}
}
\end{figure}

The ancillary 870$\mu$m and 1mm data from ALMA  provide high-resolution  imaging for our sample. We  measure the sizes of the dust continuum in the $uv$ space using 870$\mu$m and/or 1mm observations, following the same method applied in \cite{Puglisi2019} and \cite{Valentino2020CI}. 
The size measurements are listed in Table~\ref{tab:6}. The six sources are well resolved with FWHM sizes in the range of 1.7--3.4 kpc.
In Fig.~\ref{size_LIR}, we compare sizes against a control sample of MS galaxies at $z\sim1.2$ \citep{Valentino2020CI,Puglisi2019} and from the GOODS-ALMA sample (\citealt{Franco2020alma,Gomez-Guijarro2022catalog,Gomez-Guijarro2022}). 
Our sample has a dust continuum FWHM which is $2.4\pm1.3$
times smaller than that of comparable FIR luminous galaxies at $z\sim$1--2, while it has comparable FWHM sizes with the GOODS-ALMA sample that was selected from high-resolution ALMA 1mm images (resolution$=0.6''$, \citealt{Franco2020alma}).
Interestingly, an anti-correlation is seen in the figure, that is, 
\begin{equation}
   {\rm log({\rm FWHM~size /kpc})} = -0.38\times{\rm log}(L_{\rm IR}/10^{10}L_{\odot})+1.42,
\end{equation}
with 1$\sigma$ scatter of 0.24 dex, indicating galaxies with higher IR luminosity tend to have a more compact dust morphology. 
This correlation appears to be intrinsic rather than  due to selection effects: as  the local spirals, LIRGs, and ULIRGs (green crosses  in Fig.~\ref{size_LIR}) follow a similar anti-correlation, where high-z dusty galaxies have either larger sizes than the local ones, higher luminosities, or a combination there of. This is consistent with the findings of \cite{Tacconi2006}, who showed that  SMGs at $z=$2--3 have larger sizes than local ULIRGs, suggesting that high-z SMGs are scaled-up versions of local ULIRGs at the maximum starburst limit.

This correlation is in opposition with the positive correlation reported in \cite{Fujimoto2017size}. We note that the sample of \cite{Fujimoto2017size}  was selected from the ALMA archive, which mixes all types of galaxies observed with variable configurations and sensitivities.

\subsection{Near-infrared and IRAC photometry and stellar mass}

We cross-matched this sample to the deepest wide-field near-infrared(NIR)/IRAC catalogs from COSMOS2020 \citep{Weaver2021cosmos2020}. 
Two sources, ID6309 and 7549, are found in both \texttt{The Farmer} and \texttt{Classic} versions. Because of the typically greater precision of profile-fitting, we chose to adopt photometry from \texttt{The Farmer} for these two sources.
ID5340 is not included in the catalog derived using \texttt{The Farmer} because it is in a masked region where photometry was not performed. However, it is found in the \texttt{Classic} catalog but the IRAC channel 2, 3, and 4 measurements are missing, and so for this source we use \texttt{Classic} photometry (VISTA and IRAC channel 1), and IRAC channel 2 to 4 photometry from the S-COSMOS catalog \citep{Sanders2007scosmos}.
For the remaining sources in COSMOS (including the four in \citealt{Jin2019alma}), their UltraVISTA and IRAC photometry are measured by performing new measurements with the \texttt{The Farmer} \citep{Weaver2021cosmos2020}, adopting point-like models centered on their ALMA positions. Models of known sources neighboring these positions from COSMOS2020 were re-fitted to improve deblending, which is consistent with the methods of the original COSMOS2020 catalog from \texttt{The Farmer}. 
The IRAC photometry of the GOODS-N source ID12646 is from the GOODS-Spitzer project (PI: M. Dickinson).
Stellar masses, attenuation, and luminosity-weighted ages are summarized in Table~\ref{tab:5}, which are estimated using the MICHI2 code \citep{Liu_DZ2021MiChi2} by fitting the NIR+IRAC photometry with star-formation templates from BC03 \citep{BC03}, allowing for dust attenuation \citep{Calzetti2000} and multiple ages. 
The $E(B-V)$ values indicate high attenuation, that is, $2<A_{\rm V}<3,$ if adopting the Milky Way reddening parameter $R_{\rm V}=3.1$ and $3<A_{\rm V}<4.6$ if adopting $R_{\rm V}=5$ for dense dust, which is in line with the optically thick dust suggested by FIR data (Section 4). The age is not reliably constrained because of the high attenuation.

\begin{table}[ht]
{
\caption{Stellar SED fitting results}
\label{tab:5}
\renewcommand\arraystretch{1.1}
\centering
\begin{tabular}{ccccccccccc}
\hline\hline
     ID       &   log($M_*/M_\odot$)    &  $E(B-V)$ &  Age$^*$ [Gyr] \\
 \hline
3117  &    11.24$\pm$0.11     &   0.71$\pm$0.05 & 0.65$\pm$0.21 \\
5340 &   10.49$\pm$0.44  &    0.63$\pm$0.31   & 0.33$\pm$0.31\\
6309  &    10.82$\pm$0.15   &  0.77$\pm$0.06  &  0.17$\pm$0.10\\
7549 &   10.84$\pm$0.19  &   0.88$\pm$0.09  & 0.17$\pm$0.10 \\
9316 & 10.79$\pm$0.37   &   0.81$\pm$0.15  &  0.31$\pm$0.30\\
12646 &   10.99$\pm$0.35   &    0.92$\pm$0.19  & 0.29$\pm$0.30  \\
\hline
\hline
\end{tabular}\\
}
{$^*$Age is luminosity-weighted.} 
\end{table}

Note that we are fitting stellar and dust SEDs separately rather than using an energy balance between optical and FIR. This is because dust and stellar components are often found to be spatially unassociated in high-z dusty galaxies, e.g., \cite{Franco2018GOODSS,Hodge2019}. Literature work has shown that in this type of galaxy the star formation from optical emission is just a tiny fraction of the total SFR \citep{Puglisi2017,Calabro2018}. Therefore, there is no reason to assume that the dust and stellar are well mixed.

\begin{table*}[ht]
{
\caption{Physical quantities}
\label{tab:6}
\renewcommand\arraystretch{1.1}
\centering
\begin{tabular}{ccccccccccc}
\hline\hline
     ID     &  FWHM size &  $i$  &  $\Sigma_{\rm SFR}$ & $M_{\rm dyn, 2Re}$ & ${M_{\rm gas,CI}}$  &  $\tau_{100\mu m}$ & $M_{\rm gas,dyn}$ & $\tau_{\rm dep}$ \\
                 &   [$''$ (kpc)]    & [$\degree$]   &   [${\rm M_\odot yr^{-1} kpc^{-2}}$]   &  [$10^{10}$M$_\odot$]   &      [$10^{10}$M$_\odot$]       &         &  [$10^{10}$M$_\odot$] &  [Myr]   \\
 \hline
3117  & 0.28$\pm$0.03 (2.03$\pm$0.24)  & 16$\pm$41  &   96$\pm$46  & --   & --  & 3.8$\pm$1.4 &  -- &  -- \\
5340 & 0.24$\pm$0.04 (1.66$\pm$0.29) & 63$\pm$17 &   135$\pm$50  &   9.5$\pm$3.4     &15.2$\pm$2.6    & 3.8$\pm$1.5 &  5.5$\pm$5.3  &  260$\pm$58\\
6309 & 0.47$\pm$0.04 (3.26$\pm$0.25)   & 60$\pm$6 & 29$\pm$7 &  6.5$\pm$0.8   & 21.5$\pm$4.7     & 1.3$\pm$0.3 &  -- &  442$\pm$124\\
7549 & 0.49$\pm$0.02 (3.36$\pm$0.17)  &  68$\pm$5 &   28$\pm$3 &  30.2$\pm$2.5   & 28.5$\pm$6.0    & 3.5$\pm$1.0 & 22.3$\pm$14.4 & 582$\pm$160 \\
9316 & 0.44$\pm$0.03 (3.03$\pm$0.22)  & 46$\pm$11 &   70$\pm$20  &   48.5$\pm$17.8  & $17.3\pm$1.0  & 1.5$\pm$0.3    & 32.5$\pm$15.1  &  173$\pm$44\\
12646 & 0.26$\pm$0.03 (1.86$\pm$0.21)   &  32$\pm$13  &    64$\pm$18  & 23.5$\pm$17.0  & 11.8$\pm$2.5    & 2.2$\pm$0.6 & 9.0$\pm$15.7  & 331$\pm$90 \\
\hline
\hline
\end{tabular}\\
}
{Notes: $i$: inclination angle; $M_{\rm dyn, 2Re}$: dynamic mass within FWHM size; SFR surface density $\Sigma_{\rm SFR} \equiv {\rm SFR/(2\pi {R_{e}}^{2})}$, where ${\rm R_e}$ is half of the FWHM size; 
$\tau_{100\mu m}$ is the opacity at 100$\mu$m rest frame wavelength derived from $M_{\rm d,thin}$ and sizes;$M_{\rm gas,CI}$: molecular gas mass from [CI](1-0) adopting Eq. (2) in \cite{Valentino2018CI}; $M_{\rm gas,dyn}=M_{\rm dyn,2Re}\times 0.8 - M_{*}$. 
} 
\end{table*}

\begin{figure*}
\setlength{\abovecaptionskip}{-0.1cm}
\setlength{\belowcaptionskip}{-0.4cm}
\centering
\includegraphics[width=0.98\textwidth,trim={0cm 0cm 0cm 0cm}, clip]{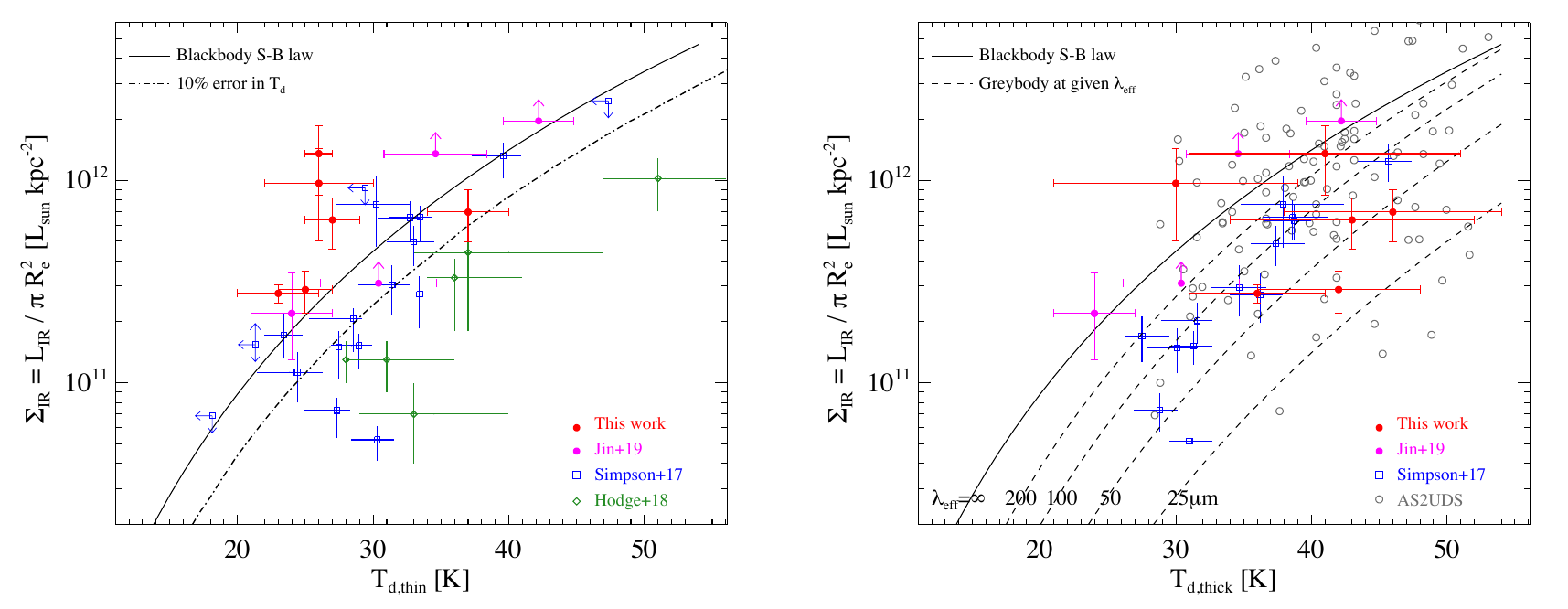}
\caption{
IR luminosity surface density vs dust temperature for dusty galaxies in the literature and this work. Left and right panels show dust temperatures from optically thin and thick models, respectively. Blue squares mark data points from \cite{Simpson2017}; red diamonds show the ALESS sample in \cite{Hodge2019}. The AS2UDS sources \citep{Stach2019AS2UDS} are shown in open circles, which are calculated using results from \citep{Gullberg2019} and \cite{Dudzeviciute2020}. The Stefan-Boltzmann (S-B) law is shown as a solid curve, representing the expected limit for a black body. The dot-dashed curve marks the 10\% systematic bias of $T_{\rm d, thin}$, where above the curve $T_{\rm d}$ would be underestimated more than $10\%$ by thin models. Dashed curves show the expected relations for general models with different $\lambda_{\rm eff}$ as labeled. 
This figure provides a diagnostic method identifying optically thick dust.
\label{Sigma_Td}
}
\end{figure*}

\section{Evidence for optically thick dust in the FIR}

\subsection{Cold dust temperature against high IR surface brightness $T_{\rm d}$--$\Sigma_{\rm IR}$}

Figure~\ref{Sigma_Td} shows that the new sample has cold dust temperatures and high IR surface densities that are more extreme than those of the other dusty galaxy samples.
By performing a bootstrapped Kolmogorov-Smirnov test in $T_{\rm d}$--$\Sigma_{\rm IR}$ space, we found a low probability of similarity (median$=2.4\times 10^{-3}$, semi-interquartile$=3.5\times 10^{-3}$ ) between this sample (including \citealt{Jin2019alma}) and that of \cite{Simpson2017}, indicating that these galaxies belong to a new population and are significantly different from general high-z submm galaxies in the literature.
Formally, these measurements are inconsistent with the Stefan-Boltzmann law even when accounting for their uncertainties and error bars. 
The Stefan-Boltzmann (S-B) law applies to black body emission, and acts as an upper envelope to the measurable SFR density in real galaxies. It has been shown empirically to confine the range of observations of previously known samples of distant dusty galaxies; see for example \citealt{Simpson2017}. 
However, the Stefan-Boltzmann law is only valid for black body emission that is optically thick at all wavelengths, and it is unclear whether or not grey bodies like galaxies could break the Stefan-Boltzmann envelope.

Here we demonstrate that the Stefan-Boltzmann law is a strict upper limit that should not be violated in any case. 
The Stefan-Boltzmann law states that the emission of a black body depends only on its size and temperature $T$. The emission is optically thick at all wavelengths, and the emitted energy is $E = A \sigma T^4$. For the case of a grey body, it has $E' = A \epsilon \sigma T^4$ where $\epsilon$ is the emissivity ($\epsilon = 1 - e^{-\tau}$, $\tau = (\nu/\nu_{0})^\beta$) , and $E' < E$ because $\epsilon<1$. We integrated the formula of grey bodies with different effective wavelengths $\lambda_{\rm eff}$ from $1\mu$m up to 1E7$\mu$m, that is, from optically thin to black body.
The resulting Stefan-Boltzmann law for grey body can be described as ${\rm log}L_{\rm IR}=a+b\times {\rm log}T_{\rm d}$ in cgs units, and the coefficients are listed in Table~\ref{tab:7} for different $\lambda_{\rm eff}$.
\begin{table}[ht]
\centering
{\caption{Grey body coefficients for ${\rm log}L_{\rm IR}=a+b\times {\rm log}T_{\rm d}$ in cgs units}
\label{tab:7}
\begin{tabular}{ccc}         
\hline\hline
  \multicolumn{1}{c}{$\lambda_{\rm eff}$ [$\mu$m]} &
  \multicolumn{1}{c}{a} &
  \multicolumn{1}{c}{b} \\
\hline
  1.0 & -10.55 & 5.78\\
  10 & -8.72 & 5.76\\
  25 & -7.92 & 5.67\\
  50 & -7.21 & 5.48\\
  100 & -6.41 & 5.16\\
  150 & -5.94 & 4.94\\
  200 & -5.64 & 4.79\\
  300 & -5.26 & 4.59\\
  400 & -5.03 & 4.47\\
  1000 & -4.54 & 4.19\\
  1.0E7 & -4.22 & 4.00\\
\hline\hline
\end{tabular}
}
\end{table}
When a galaxy is thick at all wavelengths, it emits like a black body ($L_{\rm IR}\propto T^4$) rather than a grey body ($L_{\rm IR}\propto T^{4+\beta}$ in the thin limit).

In Fig.~\ref{Sigma_Td}, we show the $\Sigma_{\rm IR}-T_{\rm d}$ for different galaxy samples and the normalized grey-body models with different $\lambda_{\rm eff}$ (dashed curves). The thicker dust models tend to approach higher IR surface density with colder $T_{\rm d}$, but they are limited at the Stefan-Boltzmann law ($\Sigma_{\rm IR}=5.5\times10^5 T_{\rm d,thin}^4$) where $\lambda_{\rm eff}$ is infinite.
The SFR densities derived from global size measurements are lower limits to the actual $\Sigma_{\rm IR}$. It is therefore noticeable that most of the galaxies in our new sample explicitly violate the Stefan-Boltzmann envelop.
In order to quantify the bias introduced by the misuse of thin models, we fit thick dust templates using thin models and compare the intrinsic temperatures to the measured ones $T_{\rm d,thin}$. The left panel of Fig.~\ref{Sigma_Td}  shows the 10\% systematic bias of $T_{\rm d,thin}$ as a dot-dashed curve; for any data points consistent with or above the curve, their $T_{\rm d}$ will be underestimated by at least 10\% by thin models. 
The expression of the dot-dash curve is $\Sigma_{\rm IR}=1.5\times10^5 T_{\rm d,thin}^{4.21}$, providing a useful limit for thin models; above this curve the systematic bias of $T_{\rm d,thin}$ becomes significant and optically thick dust is needed.
On the other hand, using $T_{\rm d}$ in optically thick models (Fig.~\ref{Sigma_Td}-right)
resolves the tension, as the galaxies show a distribution that appears comparable with what was seen for other regular dusty galaxy samples; for example, the SCUBA2 sources \citep{Simpson2017} and the ALESS sample \citep{Hodge2019}.
Therefore, the Stefan-Boltzmann law disproves the extremely cold $T_{\rm d}$ from the thin model and indicates optically thick dust.

In the right panel of Fig.~\ref{Sigma_Td} we also compare with the AS2UDS sample \citep{Stach2019AS2UDS} with quantities calculated by \cite{Gullberg2019} and \cite{Dudzeviciute2020}, finding that the AS2UDS sample has more scatter violating the Stefan-Boltzmann limit. We note that the SFR and $T_{\rm d}$ of the AS2UDS are derived using a different method from this work, namely MAGPHYS optical+FIR+radio SED fitting with optical-FIR energy balance; this could lead to systematic errors.

\subsection{High star formation efficiency from [CI](1-0)}

Dust temperature can be expressed in terms of the average intensity of the radiation field $\langle U\rangle$, according to the relation: $\langle U\rangle=(T_{\rm d}/18.9)^{6.04}$ \citep{Magdis2012SED}.
For the parameter $\langle U\rangle,$ the following relation applies $\langle U\rangle\propto{\rm SFE}/Z$ (i.e., star formation efficiency SFE over metallicity $Z$, \citealt{Magdis2012SED}). The $\langle U\rangle$ measurements for  our cold sample are approximately ten times lower than those of MS galaxies at $z=4$ from \cite{Magdis2017}, or  about 3.3 times lower than those of MS galaxies at $z\sim2$ (see Fig.~\ref{Td}-right in \citealt{Jin2019alma}).
Hence, in \cite{Jin2019alma} we discussed several interpretations for the cold dust temperatures and low $\langle U\rangle$ values: (1) the galaxies have optically thick dust in FIR, which are hot in the cores but appear cold for observers; (2) the galaxies have very low SFEs (i.e., long depletion timescales) and/or (3) the ISM is cooling efficiently due to very high metallicities.

\begin{figure}
\setlength{\abovecaptionskip}{-0.1cm}
\setlength{\belowcaptionskip}{-0.4cm}
\centering
\includegraphics[width=0.48\textwidth,trim={0cm 0cm 0cm 0cm}, clip]{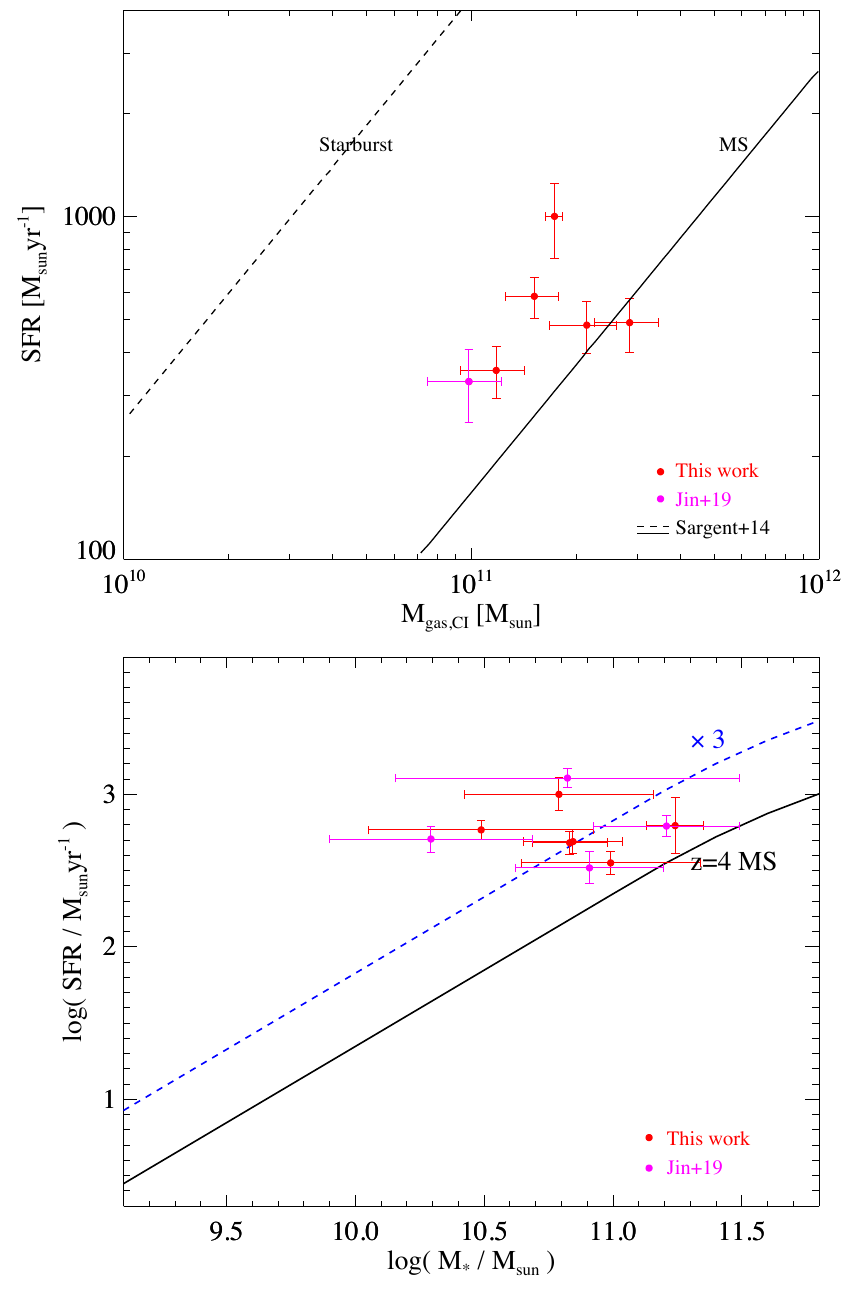}
\caption{
{\bf Top:} SFR vs molecular gas mass derived from CI. Solid and dashed lines mark the relations for MS and starbursting galaxies at $z\sim2$ in \cite{Sargent2014}. {\bf Bottom:} SFR vs stellar mass. Black line shows the MS at $z=4$ from \cite{Schreiber2015}, and the blue dashed line marks three times the MS.
Our sample galaxies show high SFEs with respect to MS galaxies and appear to be starbursts on the SFR--$M_{*}$ diagram.
\label{SFE}
}
\end{figure}

Given that five of the six sources in this work have been detected with [CI](1-0) and all FIR SEDs are well constrained, we can directly measure their SFEs to disentangle the above ${\rm SFE}-Z$  degeneracy. We estimate their molecular gas masses by adopting the empirical calibration from \cite{Valentino2018CI}, assuming a CI excitation temperature $T_{\rm exc,CI}=T_{\rm d, thick}$. SFRs are converted from IR luminosity from the FIR SED fitting, that is, ${\rm SFR}=L_{\rm IR} \times 10^{-10}$.
In Table~\ref{tab:6}, we list gas depletion times $\tau_{\rm dep}=M_{\rm mol,CI}$/SFR, the inverse of SFEs. 
We note that the IR luminosity is invariant upon optical depth, and the [CI](1-0) emission is almost always optically thin (see specific discussion in \citealt{Jin2019alma} for these cold galaxies), and therefore the measurements of gas depletion times and SFEs are robust regardless of dust optical depth. 
We show molecular gas masses and SFRs in the top panel of  Fig.~\ref{SFE}, and compare them to scaling relations for typically star-forming and starbursting galaxies at $z\sim2$ \citep{Sargent2014}. 
The two sources with the highest [CI](1-0) luminosities, ID 6309 and 7549, are in line with the MS scaling. The remaining galaxies fall between the MS and starburst scaling relations, showing SFEs higher than MS galaxies by a factor of 1.8--2.5. 
Even if adopting the canonical CI-to-gas-mass conversion from \cite{Weiss2005ci}, which scales up the $M_{\rm gas,CI}$ by a factor of two, the SFEs would
still be in line with the MS scaling.
Hence, there is no evidence for anomalously low SFE.

Metallicity might be responsible for the anomalously cold dust.
To explain these $\langle U\rangle$ measurements together with the observed SFEs, the metallicity $Z$ would need to be between six to eight times higher than that seen in MS galaxies at $z\sim2.$ This appears unrealistic for galaxies at $z\sim4$. 
However, in principle, a much higher metallicity could also affect the [CI](1-0) luminosity-to-gas-mass conversion factor, for example, CI/H$_2$ abundance scales linearly with metallicity $Z$ \citep{Heintz2020CI}. This would reduce the requirement for excess metallicity over MS galaxies to factors of two to three, but it would still be problematic for these gas-rich galaxies. Nevertheless, we note that dynamical constraints discussed in the following section rule out an anomalous [CI](1-0) luminosity-to-gas-mass conversion factor.
Therefore, the low SFE/high-$Z$ explanation for cold dust is strongly disfavored, and the optically thick dust becomes the only rational explanation left.

We note that recent studies questioned the molecular gas mass estimates from CI. For example, \cite{Izumi2020CI} found that [CI](1-0) luminosity is enhanced by a factor of ten by type 1 AGN in a local galaxy. Similar cases are also found in other local galaxies (Liu et al. 2022). Such extreme cases occur only in the very central region ($\sim$300 pc) around luminous AGN where there is non-local thermodynamic equilibrium (non-LTE). However, such bias would be diluted in unresolved galaxies or starbursting cores, significantly reducing the impact on global molecular gas estimates. This is also supported by a study of CI+CO in a local starburst galaxy by \cite{Salak2019CI}, who found [CI](1-0)/CO(2–1) is uniform within statistical errors in the central 1~kpc, and suggested that [CI](1-0) luminosity can be used as a CO-equivalent tracer of molecular gas mass.
More recently, \cite{Papadopoulos2022CI} studied line ratios of the two CI transitions of 106 galaxies and found their average excitation conditions to be strongly subthermal. This non-LTE excitation of the CI lines can add uncertainty in CI-based molecular gas mass estimates, but is only effective when the [CI](2-1) line is used, while the [CI](1-0)-based gas mass is  substantially less affected \citep{Papadopoulos2022CI}. 
Therefore, [CI](1-0) is still a good gas tracer for galaxies in this work, as supported by the dynamical estimates in the following sections.

In the bottom panel of Fig.~\ref{SFE}, we compare this sample to the MS at $z=4$ \citep{Schreiber2015}. Intriguingly, all these cold galaxies have high sSFRs (2--30$\times$ MS), and therefore clearly belong predominantly to the starburst population.

\subsection{Gas masses from optically thin dust models are overestimated}

\cite{Jin2019alma} argued that the dust mass of the cold galaxy ID85001674 is overestimated when derived using optically thin dust models, as it leads to an abnormally high gas mass that cannot be reconciled with the gas mass derived from [CI](1-0) emission and from dynamics.
On the other hand, the dust mass would be three times smaller if the dust were optically thick in the FIR and the appropriate thick models were used, bringing the gas mass into agreement with the gas mass derived from [CI](1-0) and from dynamics.
Here the same test can be performed for a larger sample thanks to the larger numbers of [CI](1-0) detections and resolved submm sizes (e.g., \citealt{Jiao2017CI,Jiao2019CI,Jiao2021CI}). We adopt the same approach for this sample, and estimate gas mass in three ways: 

(1) The gas mass $M_{\rm gas,dust}$ is derived from dust mass assuming a gas-to-dust ratio, G/D, depending on metallicity. The dust masses are derived from SED fitting (Table~\ref{tab:4}), for both the thick and thin dust cases. For G/D, the metallicities are derived using the  FMR relation \cite{Magdis2012SED}. We added a 20\% uncertainty to the derived gas mass estimates to account for variations in the assumed metallicity (see \citealt{Jin2019alma} for examples of the effect).

(2) The gas mass $M_{\rm gas,dyn}$ is derived from the dynamical masses of the galaxies, which are based on CO line widths and submm ALMA sizes, following \cite{Daddi2010SFL} and \cite{Coogan2018}; see Table~\ref{tab:6}. The dynamical masses are converted to total gas mass using the relation: $M_{\rm gas,dyn}=M_{\rm dyn,2Re}\times 0.8 - M_{*}$. The factor 0.8 is used to account for dark matter \citep{Daddi2010SFL}. 
{The inclinations are measured from the sizes of major and minor axes, which are obtained by fitting elliptical models to the ALMA emission of the galaxies}\footnote{We note that ID3117 and 6309 have relatively narrow line widths, their dynamical masses appear lower than their stellar masses. This is because ID3117 has a low and uncertain inclination angle of $i=16\pm40\degree$, making it impossible to obtain a reasonable constraint on its $M_{\rm gas,dyn}$. ID6309 has a reliable inclination measurement of $i=60\pm6\degree$ but shows excessive stellar mass with respect to the dynamical mass $M_{*}>M_{\rm dyn,2Re}$. Therefore we do not constrain their $M_{\rm gas,dyn}$.}.
The statistical dispersion in the average is propagated to the dynamical mass error.

(3) The gas mass $M_{\rm gas,CI}$ is derived from [CI](1-0) emission. To account for the small dependence on CI excitation temperature, we assume that CI excitation temperature equals to the dust temperature estimated from thick models \citep{Cortzen2020GN20}. The molecular gas mass is derived using Eq. 2 in \cite{Valentino2018CI}, as listed in Table~\ref{tab:6}. 
We note that high-J CO transitions are impractical for constraining molecular gas mass due to uncertain excitation \citep{Liudz2015,Daddi2015} and can be significantly suppressed by thick dust \citep{Papadopoulos2010Arp220}. The [CI](1-0) transition is optically thin in all reasonable conditions, and is therefore an optimal tracer of molecular gas.

\begin{figure*}
\setlength{\abovecaptionskip}{-0.1cm}
\setlength{\belowcaptionskip}{-0.4cm}
\centering
\includegraphics[width=0.98\textwidth,trim={0cm 0cm 0cm 0cm}, clip]{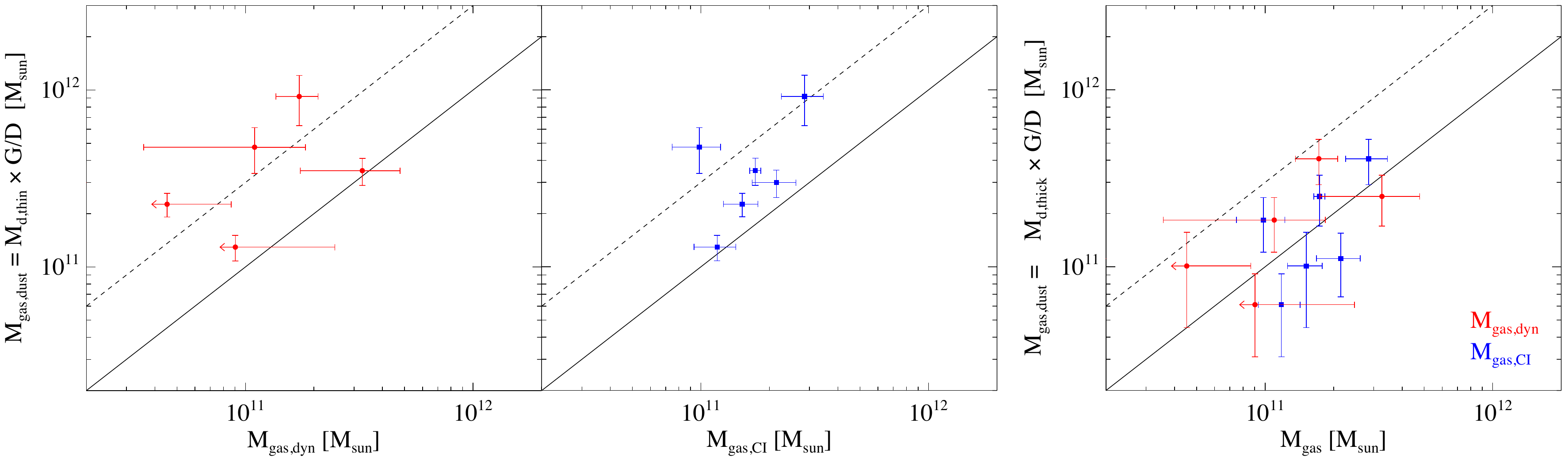}
\caption{
Gas mass derived from dust mass vs gas mass for this sample and the $z=3.62$ source in \cite{Jin2019alma}. Left panels: Comparison of gas mass from optically thin dust mass to gas mass from dynamical and [CI] estimations ($M_{\rm gas,dyn}$, $M_{\rm gas,CI}$ ), respectively. Right panel: Gas mass derived from optically thick dust to $M_{\rm gas,dyn}$ (red dots) and $M_{\rm gas,CI}$ (blue squares). Solid line shows the 1:1 identity relation and the dashed line is three times the identity line. 
Optically thin dust models significantly and systematically overestimate the dust mass of galaxies in this work, indicating that they have optically thick dust.
\label{Mgas}
}
\end{figure*}

In Fig.~\ref{Mgas}, we compare the three gas mass derivations. The gas masses from optically thin dust, $M_{\rm gas,dust}$, are explicitly (3 to 5 times) higher than those derived from dynamical mass $M_{\rm gas,dyn}$ for three galaxies, and 1.4--4 times higher than gas mass  from [CI](1-0) $M_{\rm gas,CI}$ for five galaxies.
In the left panel of Fig.~\ref{Mgas}, only the less extreme galaxy ID9316 falls well on the identity line of $M_{\rm gas,dust}-M_{\rm gas,dyn}$, while the other source ID12646 appears on the identity line but with large uncertainty on $M_{\rm gas,dyn}$. 
In terms of $M_{\rm gas,dust}-M_{\rm gas,CI}$ comparison, ID12646 is the only with $M_{\rm gas,dust}$ agreeing well with $M_{\rm gas,CI}$.
Both  comparisons indicate that the dust masses from optically thin dust models are overestimated by factors of 2--3 on average.
On the other hand, optically thick models predict dust masses that are indeed lower than the thin ones by just about the same factors of 2--3 (Fig.~\ref{Mgas}-right) and produce estimates that are fully consistent on average with both CI and dynamical derivations.

We note that the dynamical masses are calculated using high-J CO(4-3) velocity widths, while  sizes are from 1mm dust continuum; that is, we are assuming that high-J CO gas and dust emission are equivalent. Although in some bright SMGs the low-J CO(1-0) sizes  can be much larger than the high-J ones (e.g., \citealt{Ivison2011CO,Riechers2011CO10}), 
the dust continuum sizes in typical distant star-forming galaxies  are found to be consistent with the high-J CO sizes  \citep{Puglisi2019},  and the high-J CO accurately  reflects the star forming gas \citep{Daddi2015,Liudz2015} that emits from the same region as the dust.
We were only able to verify this equivalence for the ALMA 3mm-detected ID9316 in our sample, and find that its CO(4-3) size ($0.39''\pm0.07''$) is in agreement with the dust continuum size (Table 6).

 \subsection{High dust opacity at FIR}

\begin{figure}
\setlength{\abovecaptionskip}{-0.1cm}
\setlength{\belowcaptionskip}{-0.4cm}
\centering
\includegraphics[width=0.48\textwidth]{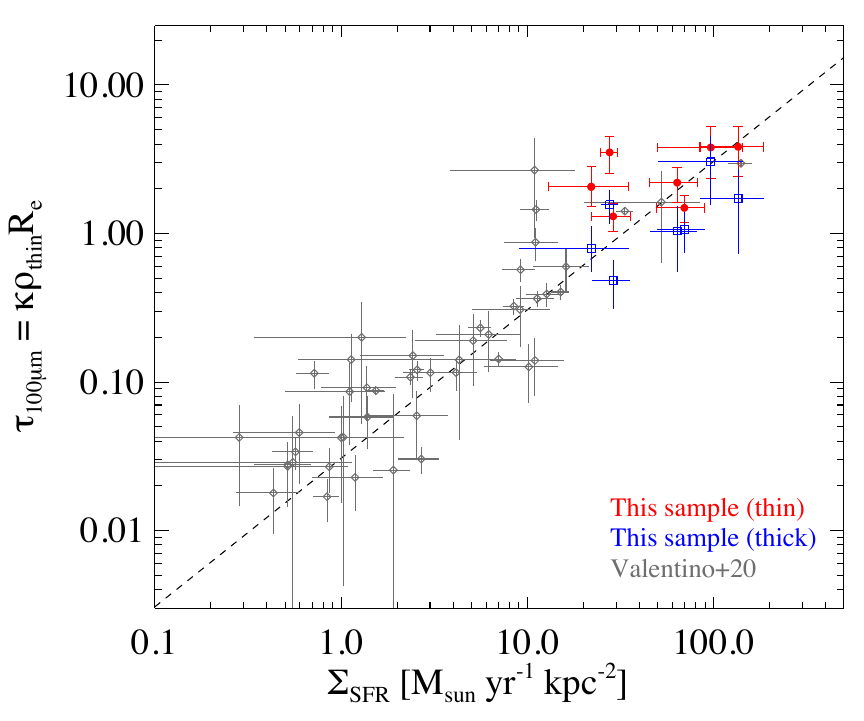}
\caption{%
Dust opacity $\tau_{100\mu m}$ vs SFR surface density $\Sigma_{\rm SFR}$. The dashed line shows a fit to the data points under the optically thin assumption, i.e., $\tau_{\rm 100\mu m}=3.1\times10^{-2}\Sigma_{\rm SFR}$. 
In both the optically thin and thick models, our sample shows high dust opacity at FIR wavelengths.
\label{tau_sigma}
}
\end{figure}

Figure~\ref{tau_sigma}  shows direct estimates for the dust opacity at 100$\mu$m rest-frame,that is, $\tau_{\rm 100\mu m}=\kappa\rho R_{e}$, where for $\kappa$ we are adopting the value from \cite{Jones2013dust} and $\rho$ is the dust volume density assuming spherical geometry. Similarly to what was discussed in terms of the Stefan-Boltzmann law, because the galaxy might actually be composed of clouds, the opacity derived in this way is an upper limit.
For our sample, we show $\tau_{\rm 100\mu m}$ derived for both optically thin and thick cases.
 We directly adopted the dust masses (optically thin) and sizes from \cite{Valentino2020CO}, as a comparison sample. These were measured in a way that is consistent with the method used in this work.  
As shown in Fig.~\ref{tau_sigma}, a correlation is seen between $\tau_{\rm 100\mu m}$ and $\Sigma_{\rm SFR}$ following $\tau_{\rm 100\mu m}=3\times10^{-2}\Sigma_{\rm SFR}$. This correlation is in fact probably driven by the correlation between dust mass and IR luminosity. 
 
Remarkably, the $\tau_{\rm 100\mu m,thin}$ of our galaxies are all greater than 1, even for the optically thick case, which already suggests that the FIR emission at 100$\mu$m is generally optically thick, and sometimes very much so. 
In the optically thick case, as shown by the blue squares, the opacity is systematically lower simply because the dust masses are lower, but it is still high ($0.5<\tau_{\rm 100\mu m}<3.0$ with a median of 1.1). This further demonstrates the need for optically thick dust in these galaxies.

\section{Discussion}

\subsection{Caution for gas mass estimates based on dust continuum}

Nowadays, as an alternative approach to CO observations, the Rayleigh-Jeans dust continuum emission, is  widely used to estimate dust and molecular ISM masses at high redshift \citep{Eales2012,Magdis2012SED,Scoville2016,Scoville2017ISM}. 
However, dust emission is presumed to be optically thin in literature studies, and calibrated relations are based on this hypothesis. 

This assumption will work for typical galaxy populations, but would become problematic for dusty galaxies with compact optically thick dust similar to those of our sample. As already discussed in Sect. 4, their $M_{\rm dust}$, $M_{\rm gas}$, and SFEs can be misjudged by the simple Rayleigh-Jeans flux-to-gas conversion if using optically thin fits. The dust masses in this work are overestimated by $2.2\pm0.5$ using optically thin fits, and simulations indicate that such overestimation can even be up to a factor of 20 when accounting for variable emissivity \citep{Fanciullo2020}. 
Therefore, conventional gas conversion based on dust continuum should be  treated with caution particularly for the most luminous systems.

Moreover, the Rayleigh-Jeans slopes of cold dusty galaxies are more substantially impacted by the CMB at high-z, which further complicates the dust and gas mass estimates. 
According to the SED results with and without CMB (the red and black curves in Fig.~\ref{sed}), the 3mm continua can be lower than the intrinsic fluxes by a factor of 1.3--1.6. This indicates that models that do not account for the impact of CMB would bias the dust and gas masses in apparently cold galaxies, this time towards underestimates. 
Several literature studies are using 870$\mu$m and 1mm continuum observations to constrain gas masses of galaxies at intermediate redshifts for convenience. It is true that the CMB impact is much smaller at these wavelengths, but the dust mass measurements are less accurate closer to the peak of the SED.
Going to longer wavelengths like 2mm or 3mm can provide potentially more reliable derivations, but this only works when properly accounting for the CMB.

\subsection{A procedure to identify optically thick sources}

Based on our results, we suggest using the following procedure in order to check whether optically thick models should be considered for a given source or sample.

(1) Calculate the dust opacity (upper limit) $\tau_{\rm 100\mu m}$ and SFR surface density $\Sigma_{\rm SFR}$ (Fig.~9). The threshold above which one should seriously consider optically thick dust is $\tau_{\rm 100\mu m}\gtrsim1$ and/or $\Sigma_{\rm SFR}\gtrsim20~M_{\odot}~{\rm yr^{-1}~kpc^{-2}}$ .

2) If a line width measurement is available (e.g., from CO), check the dynamical estimate of gas masses and compare them with gas masses derived from dust using optically thin models. If the latter have substantially larger estimates than the former, this points towards optically thick IR dust. 
If CI is available, do the same with gas estimates from CI. We note that CI gas estimation also depends on CI/H2 abundance. 

3) If a dust temperature $T_{\rm d}$ can be measured from the SED, compute its location in a $\Sigma_{\rm IR}-T_{\rm d}$ diagram (Fig.~6) and compare this to the {Stefan-Boltzmann} limit. Optically thick dust is needed if $\Sigma_{\rm IR}\gtrsim1.5\times10^5 T_{\rm d,thin}^{4.21}$(10\% bias limit) and is indispensable for galaxies violating the {Stefan-Boltzmann} law. Also, one can compare to the $T_{\rm d}$ versus redshift trend for normal galaxies.

4) If a [CI] excitation temperature is available, a significantly higher value than $T_{\rm d,thin}$ from optically thin models will also suggest optically thick dust (e.g., \citealt{Cortzen2020GN20}).

\subsection{The abundance of optically thick and compact dusty galaxies}

Given the relevant properties of the IR optically thick and compact dusty galaxy population discussed in this work, an important question to tackle is how common these galaxies are. Fully answering this question would require much more extensive follow-up spectroscopy on significantly larger samples. Meanwhile, we can attempt some estimate using the larger photometric samples. 

One interesting aspect is  
their fraction among the general population of FIR-bright galaxies, for example  at $z\sim4$. We can estimate this  using the  super-deblended catalogs in COSMOS and GOODS-North \citep{Jin2018cosmos,Liu2018}. 
Adopting the equivalent IR luminosity limit that characterises this sample $L_{\rm IR}>4\times10^{12}L_{\odot}$, 
there are 154 (COSMOS) and 19 (GOODS-N) galaxies found with NIR/optical photometric redshift  $3.5<z<4.5$, respectively. We assume conservatively that all these NIR-detected sources are optically thin  (and thus at the correct redshift). This places a strict limit on  the fraction of thick dust cases   $\gtrsim4$\% ($>$7/161) in COSMOS and $\gtrsim10$\% ($>$2/21) in GOODS-North, which are roughly consistent given the small-number statistics.

This is a strict lower limit, particularly because our spectroscopic sample of optically thick galaxies at $z\sim4$ is not complete in absolute terms. The  majority (78\%) of the $z>6$ candidates that we followed up were actually found to be $z\sim4$. 
\cite{Jin2018cosmos} presented a sample of 85 FIR-detected candidates at $z>4$--7. A similar fraction of those $z_{\rm phot,FIR}>6$ candidates would be expected at lower redshift, with their $z_{\rm phot,FIR}$ being overestimated because of thick dust.  
Counting  only those $z>6$ candidates with $L_{\rm IR}>6\times10^{12}L_{\odot}$ would be  consistent with our luminosity limit $L_{\rm IR}>4\times10^{12}L_{\odot}$ if they were instead at $z\sim4$; we can add up to 30 sources to the sample of optically thick galaxies at $3.5<z<4.5$. This would imply a more realistic fraction of 20\%  IR-luminous sources at $z\sim4$ that are optically thick in their dust emission. This is similar to the thick fraction that we estimated from the \cite{Simpson2017} sample, applying the proposed Stefan-Boltzmann limit in Fig.~\ref{Sigma_Td}.

In terms of comoving density, we have to take into account that the completeness of the super-deblended catalogs  is  differentially affected by blending, meaning that effectively deep limits are only reachable in those regions where crowding is less severe than average. Using an approximate reduction of $2-3$ in the effective volume, our source counts imply a number density of order 0.5$\times10^{-5}$~Mpc$^{-3}$.  Their contribution to the cosmic SFRD would be of order 2.5$\times10^{-3}M_\odot$~yr$^{-1}$~Mpc$^{-3}$. This is comparable to  the contribution of optically dark galaxies \citep{Wang2019Nature}, which are also $z\sim4$ candidates; although these remain largely unconfirmed with spectroscopy.

\section{Conclusions}

Using NOEMA and ALMA observations, we spectroscopically confirmed redshifts for six dusty galaxies at $z\sim4$, solidly detecting [CI](1-0) and CO transitions. We investigated the nature of these galaxies together with the similar sample of four analogous objects from Jin et al. 2019. Our conclusions are as follows:

1. These galaxies show cold dust temperatures, compact morphology, and abnormally steep Rayleigh-Jeans slopes.  This work confirms the presence of a galaxy population at $z\sim4$ with seemingly cold temperatures.

2. In all "cold" galaxies, we detect the impact of the CMB on their mm observables. The CMB reduces their 3mm continuum observed fluxes and results in steep Rayleigh-Jeans slopes. 

3. This sample shows compact dust morphology. Combining with literature data, we find a negative correlation between dust continuum size and IR luminosity.

4. We provide multiple pieces of evidence for optically thick dust in these galaxies: (1) abnormally cold dust temperature from optically thin models with respect to high IR surface brightness, violating the Stefan-Boltzmann law; (2) they have higher SFEs than MS galaxies at $z\sim2$, which disfavors the mechanism of low-efficiency star formation with rapid metal enrichment, and they will evolve to massive quiescent galaxies if there is no external gas supply; (3) the high gas mass from the thin dust model shows explicit discrepancy with the gas mass derived from CO gas dynamics and [CI](1-0) emission, indicating that dust mass is overestimated with thin dust models; (4) high optical opacity at FIR after accounting for optically thick dust mass. 
These taken together confirm the presence of optically thick dust in this sample, paving the way for selection of large samples of optically thick dust in the early Universe.

5. We roughly estimated the abundance of galaxies with optically thick dust, which consists of a substantial population (at least 10\%) of dusty star-forming galaxies with $L_{\rm IR}>4\times10^{12}~L_{\odot}$ at $z\sim4$.

Why these dusty star-forming galaxies are so compact (and therefore optically thick in FIR) is still an open question. The high dust compactness in lower redshift samples has been ascribed to mergers (e.g., as in \citealt{Puglisi2021b}), gas inflow, or auto-regulation from star formation compression \citep{Gomez-Guijarro2022}. Further follow-up studies are needed to clarify the nature of compact, dusty star-forming galaxies through the highest redshifts.

\begin{acknowledgements}
This work is based on observations carried out under projects number S18DI, W20DM, W17EK, W18EY and W19DQ with the IRAM Interferometer NOEMA. IRAM is supported by INSU/CNRS (France), MPG (Germany) and IGN (Spain).
This paper makes use of the following ALMA data: ADS/JAO.ALMA 2013.1.00118.S, 2016.1.00279.S, 2016.1.00463.S, 2016.1.00478.S and 2018.1.00874.S. 
ALMA is a partnership of ESO (representing its member states), NSF (USA), and NINS (Japan), together with NRC (Canada), MOST and ASIAA (Taiwan), and KASI (Republic of Korea), in cooperation with the Republic of Chile. The Joint ALMA Observatory is operated by ESO, AUI/NRAO, and NAOJ.
SJ is supported by the European Union's Horizon Europe research and innovation program under the Marie Sk\l{}odowska-Curie grant agreement No. 101060888.
The Cosmic Dawn Center (DAWN) is funded by the Danish National Research Foundation under grant No. 140.
GEM, SJ and DBS acknowledge the Villum Fonden research grants 37440, 13160.
JRW acknowledges support from the European Research Council (ERC) Consolidator Grant funding scheme (project ConTExt, grant No. 648179).
FV acknowledges support from the Carlsberg Foundation Research Grant CF18-0388 ``Galaxies: Rise and Death".
YG acknowledges the National Key R\&D Program of China No. 2017YFA0402704 and National Natural Science Foundation of China grant Nos. 11861131007 and 12033004.
QSG is supported by the National Key Research and Development Program of China (No. 2017YFA0402703), and by the National Natural Science Foundation of China (No. 11733002, 12121003, 12192220 and 12192222), and by the science research grants from the China Manned Space Project with NO. CMS-CSST-2021-A05.
\end{acknowledgements}

\bibliographystyle{aa}
\bibliography{biblio}

\end{document}